\documentclass[journal]{IEEEtran}
\IEEEoverridecommandlockouts
\usepackage{cite}
\usepackage[pdftex]{graphicx}
\usepackage{caption}
\usepackage{epstopdf}
\usepackage{dblfloatfix}
\usepackage{blindtext}
\usepackage{float}
\usepackage[cmex10]{amsmath}
\usepackage{amssymb}
\usepackage{color}
\usepackage{bm}
\usepackage{amsfonts}
\usepackage{amsmath,mathtools,amssymb}
\usepackage{mathrsfs}
\usepackage{algorithm}
\usepackage{algpseudocode}
\usepackage[mathlines]{lineno}
\usepackage{tablefootnote}
\usepackage[singlelinecheck=false]{caption}

\usepackage{amsmath,mathtools,amssymb}
\usepackage{diagbox}

\DeclarePairedDelimiter{\norm}{\lVert}{\rVert}
\newcommand{\normmm}[1]{{\left\vert\kern-0.25ex\left\vert\kern-0.25ex\left\vert #1 
   \right\vert\kern-0.25ex\right\vert\kern-0.25ex\right\vert}}

\newcommand{\bb}[1]{#1}

\begin{document}

\title{Learned Trimmed-Ridge Regression for Channel Estimation in Millimeter-Wave Massive MIMO}

%
%
        
\author{Pengxia~Wu,
        Julian~Cheng,~\IEEEmembership{Fellow,~IEEE},
		Yonina C. Eldar,~\IEEEmembership{Fellow,~IEEE},\\
		and John M. Cioffi,~\IEEEmembership{Life Fellow,~IEEE,
       }

	\thanks{P. Wu was with the School of Engineering, The University of British Columbia, Kelowna, BC V1X 1V7, Canada (email: pengxia.wu@alumni.ubc.ca).} 
	\thanks{J. Cheng is with the School of Engineering, The University of British Columbia, Kelowna, BC V1X 1V7, Canada (email: julian.cheng@ubc.ca).}
	\thanks{Yonina C. Eldar is with the Math and CS Faculty, Weizmann Institute of Science, Rehovot 7610001, Israel (e-mail: yonina.eldar@weizmann.ac.il).}
	\thanks{John M. Cioffi is with the Department of Electrical Engineering, Stanford University, Stanford, CA 94305 USA (e-mail: cioffi@stanford.edu).}
}

\maketitle
\begin{abstract}
Channel estimation poses significant challenges in millimeter-wave massive multiple-input multiple-output systems, especially when the base station has fewer radio-frequency chains than antennas. To address this challenge, one promising solution exploits the beamspace channel sparsity to reconstruct full-dimensional channels from incomplete measurements. This paper presents a model-based deep learning method to reconstruct sparse, as well as approximately sparse, vectors fast and accurately. To implement this method, we propose a trimmed-ridge regression that transforms the sparse-reconstruction problem into a least-squares problem regularized by a nonconvex penalty term, and then derive an iterative solution. We then unfold the iterations into a deep network that can be implemented in online applications to realize real-time computations. To this end, an unfolded trimmed-ridge regression model is constructed using a structural configuration to reduce computational complexity and a model ensemble strategy to improve accuracy. \bb{Compared with other state-of-the-art deep learning models, the proposed learning scheme achieves better accuracy and supports higher downlink sum rates.}

\end{abstract}

\IEEEpeerreviewmaketitle


\section{Introduction}
Millimeter Wave (mmWave) massive multiple-input multiple-output (MIMO) has become a key technology in wireless communications by offering a high data rate and spectrum efficiency for wide frequency bands and large antenna arrays \cite{heath2016an}. 
Accurate channel state information (CSI) is crucial for achieving the expected array gains, as beamforming relies on CSI to direct narrow beams precisely towards the intended users \cite{Busari2018millimeter}. To reduce the hardware and energy cost, current mmWave massive MIMO systems often adopt a hybrid analog digital (AD) architecture having a few radio frequency (RF) chains whose number is less than the number of antennas \cite{Busari2018millimeter}. Under such an architecture, conventional channel estimation schemes can be time-costly and computational-intensive because they require many pilots to acquire enough channel observations. 
On the other hand, if the number of pilots is insufficient, conventional channel estimation schemes often fail to estimate channels accurately. By contrast, beamspace channel estimation methods are desirable when the angular-domain (beamspace) channels are approximately sparse and thus can be reconstructed from a few pilot observations \cite{Berger2010application}.

Existing beamspace channel estimation approaches can be categorized into compressive-sensing-based and deep-learning-based schemes. 
The compressive-sensing-based techniques regard the pilot training and channel reconstruction process as a compressive-sensing process and reconstruct the beamspace channels using compressive-sensing algorithms \cite{yonina2012CS, yonina2015sampling}.
Many compressive-sensing-based channel-estimation schemes were proposed for beamspace channel estimation in massive MIMO \cite{gao2015spatially, gao2016structured,eltayeb2014compressive,choi2015downlink,tseng2016enhanced,
shen2016joint, rao2014distributed,liu2017closed, wu2021nonconvex} and mmWave massive MIMO systems \cite{gao2017reliable,gao2016channel_letter,Mo2018}, including methods developed to reduce feedback overhead \cite{eltayeb2014compressive, tseng2016enhanced, liu2017closed}.
\bb{Greedy algorithms such as OMP \cite{tropp2007signal} were used for massive MIMO channel estimation in \cite{Venugopal2017, coma2018channel, shen2016joint}. Some OMP variants exploited the common-support feature between subcarriers or different users \cite{rod2018frequency, gao2016channel_letter, rao2014distributed}. 
Sparse Bayesian learning (SBL) was explored for massive MIMO channel estimation and several variants exploited channel temporary correlations and common-sparsity structures in different subcarriers \cite{Srivastava2019quasi, wang2017sparse}. 
An online Bayesian learning algorithm was proposed to relieve the computational burden \cite{joseph2017noniterative}.
The approximate message passing (AMP) algorithm \cite{Donoho2009message} was also used for massive MIMO channel estimation.
Many of these existing methods rely on the common-support sparsity structure that exist in different subcarrier channels or different users' channels. 
Their performance degrades in practice when the assumptions of channel sparsity or common-support sparsity structure are damaged by power leakage and beam squint effects \cite{gao2023deeplearning, chen2021hybrid}.}

Deep-learning-based schemes are data-driven and establish models based on empirical experience and then use massive data to train the model and obtain an approximate solution. 
Many deep-learning-based channel estimation schemes have shown satisfactory performance using general-purpose deep-network models \cite{he2018deep, chun2019deep, wen2018deep, wang2018deep, yang2019deep, Ma2020data, Mashhadi2021pruning}, including the fully-connected deep neural network (DNN), the convolutional neural network (CNN), and the recurrent neural network (RNN). 
However, one major shortcoming of the deep-learning-based approach is its `black box' nature, and as a result, the developed channel-estimation schemes are difficult to interpret and analyze. Also, the built models often have large size and require massive training data and extensive parameter tuning by trial and error.

Model-based deep-learning schemes have drawn increasing attention because they leverage the well-established theoretical foundations---such as a proper system model, the known signal statistics, or existing estimation and detection algorithms---into designing a more efficient deep-learning network\cite{Shlezinger2022model, Shlezinger2023model}. The model-based deep-learning models not only result in reduced sizes, better interpretation, and smaller training data sets, they also achieve higher accuracy and faster convergence \cite{he2019model}. 
Deep unfolding is one such promising method that transforms a traditional iterative algorithm into a trainable deep-learning network \cite{hershey2014deep, Monga2021algorithm}.
Various deep unfolding algorithms were developed for sparse reconstruction, such as the learned iterative shrinkage-thresholding algorithm (LISTA) \cite{Gregor2010Learning} and the learned denoising-based approximate message passing network (LDAMP) \cite{metzler2017learned}. Variant algorithms were developed for CSI acquisition applications, such as the LDAMP and learned AMP (LAMP) \cite{Ma2021model} for beamspace channel estimation \cite{he2018deep, Ma2021model} and the basis-pursuit autoencoder for pilot matrix design \cite{wu2022deep}.

This work proposes a deep unfolding algorithm to provide fast and accurate beamspace channel estimation for mmWave massive MIMO. The top-$\ell_{K,2}$ norm \cite{Tono2017efficient} is integrated into deep-unfolding networks to promote reconstruction sparsity and accuracy. 
To design such an unfolding network, we first leverage the top-$\ell_{K,2}$ norm to derive an exact sparse reconstruction solution which works perfectly with noiseless measurements and known sparsity level. However, implementing such an iterative algorithm is challenging in practice because it has high computational complexity and degraded performance when the sparsity level is unknown and the measurements are noisy. 
To address the challenges in practical implementation, we use deep unfolding to transfer these iterations into a more robust and lower-computational learnable algorithm. The proposed deep unfolding algorithm does not rely on a Gaussian statistical prior as in SBL algorithms, nor the common-support structural sparsity assumptions as in block-sparse-recovery algorithms. Instead, the proposed method treats beamspace channels as generic sparse vectors with unknown sparsity levels, and then specifies the top-$\ell_{K,2}$ norm parameters to promote different sparsity levels in different models, and finally leverages model-averaging ensemble to boost the reconstruction accuracy.
Our contributions are summarized as follows:

\begin{itemize}

\item The trimmed-ridge regression is proposed to formulate the sparse-reconstruction problem, leading to more accurate solutions than the Lasso and ridge regression. An iterative algorithm is derived to reconstruct exact-sparse samples precisely.

\item A deep unfolding network is proposed to improve the robustness and reduce the computational time in practical applications.  With an interpretable structure of only a few layers, the constructed deep network can reconstruct beamspace channels faster and more accurately than the iteraive algorithms.

\item A specific configuration is proposed to reduce computational complexity without degrading the model-prediction accuracy. A model ensemble strategy is proposed to improve model-prediction accuracy. Numerical results confirm the superior performance of the proposed deep-learning scheme in terms of both channel-estimation error and downlink sum rate.  
 
\end{itemize}

This paper is organized into six sections. Section II formulates the problem of channel estimation in a mmWave massive MIMO system. Section III introduces the iterative trimmed regression algorithm, which is further expanded upon in Section IV, where the algorithm evolves into the learned trimmed ridge regression variant. Section V is dedicated to the exposition of numerical findings, while Section VI encapsulates conclusions.


\section{System Model}
\subsection{Channel Estimation in a MmWave Massive MIMO System} \label{II_A}

\begin{figure}[!t]
\centering
\includegraphics[scale=0.5]{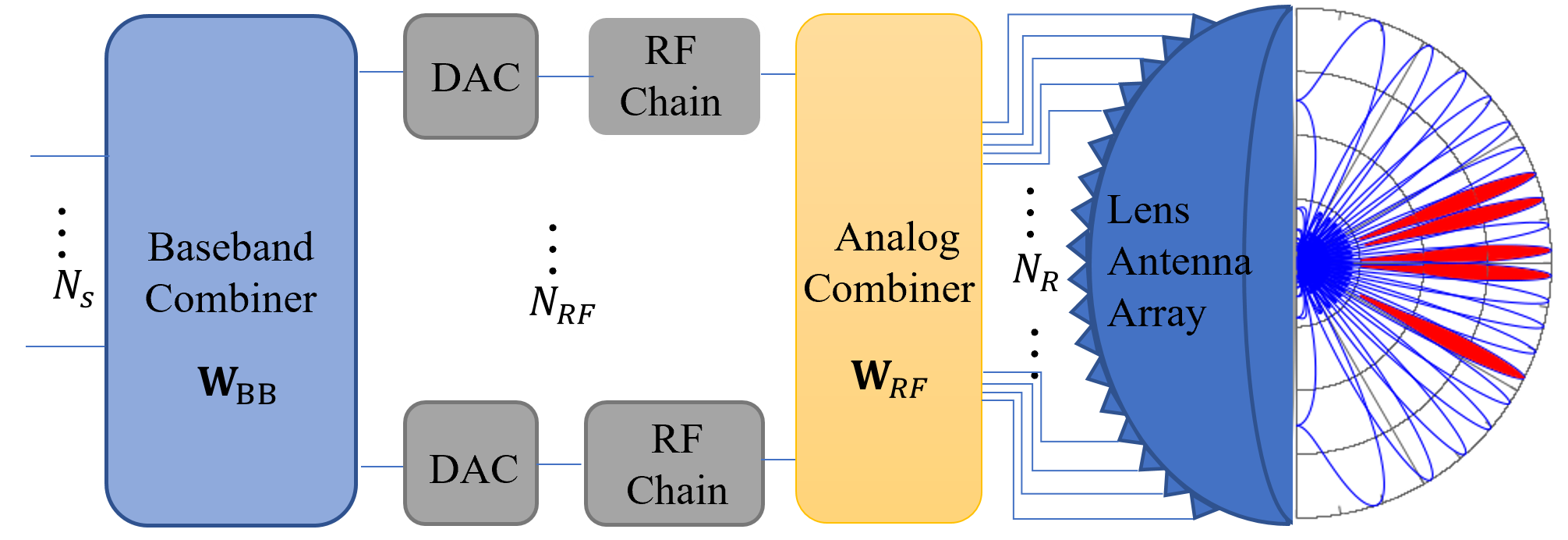}
\captionsetup{justification=centering}
\caption{A mmWave massive MIMO system with a lens antenna array.}
\label{NMSE_Noise_EmbDim128}
\end{figure}

The system model is a mmWave massive MIMO system operating in time division duplex mode. 
As in Fig. 1, the base station (BS) has a lens antenna array with $N$ antennas and $N_{RF}$ RF chains, where $N_{RF} \ll N$. 
The BS serves $U$ users and each user has a single antenna. 
For the uplink transmission, the model adopts a hybrid combiner as $\mathbf W = \mathbf W_{BB} \mathbf W_{RF} \in \mathbb{C}^{N_{S}\times N}$, where $\mathbf W_{BB} \in \mathbb{C}^{N_{S} \times N_{RF}}$ is the digital baseband combining matrix, $\mathbf W_{RF} \in \mathbb{C}^{N_{RF} \times N}$ is the analog combining matrix, and $N_{S}$ denotes the number of independent data streams. 
The model includes the case where $N_{S}=N_{RF}$ and then sets $\mathbf W_{BB}=\mathbf I_{N_{RF}}$; thus, the combiner becomes $\mathbf W = \mathbf W_{RF} \in \mathbb{C}^{N_{RF}\times N}$, where the analog combiner network implements one-bit phase shifters. 
To estimate channels for all $U$ users, each user repeatedly transmits a known pilot sequence $\mathbf p_u \in \mathbb{C}^U$ to the BS across $B$ time blocks. 
Without loss of generality, the received pilots for a user at the $m$th time block are 
\begin{equation} \label{overline_r_m}
\begin{aligned}
\mathbf r_{m} 
& =  \mathbf W_{m} \sum \limits_{u=1}^U \mathbf h_u \mathbf p_u^T \mathbf p_u^* + \mathbf n_{m, u} \\
& = \mathbf W_m \mathbf h_u + \mathbf n_{m, u},
\end{aligned}
\end{equation} 
where $\mathbf r_{m}$ is the pilot observation for the a user at the $m$th time block, $\mathbf h$ is the channel vector, and $ \mathbf n_{m}$ is the effective noise vector.

For narrowband block fading channels, a \mbox{spatial-domain} channel can be expressed as\cite{gao2017reliable} 
	\begin{equation}\label{h_u}
	\mathbf h = \sqrt{\frac{N_{BS}}{N_p}} \sum _{l=1} ^{N_p} \beta^{(l)} \bm \alpha (\phi ^{(l)}), 
	\end{equation}
where $N_p$ is the number of scattering clusters, $l = 1$ is the index for the line-of-sight path, $2 \leq l \leq N_p$ is the index for non-line-of-sight (NLoS) paths, $\beta^{(l)}$ is the complex path gain, and $\bm \alpha (\phi  ^{(l)})$ is the array steering vector at the spatial direction $\phi ^{(l)}$ for the $l$th path.
The array steering vector contains a list of complex spatial sinusoids representing the relative phase shifts of the incident far-field waveform across the array elements. For an $N$-element uniform linear array, $ \bm \alpha (\phi ^{(l)})$ is 
    \begin{equation}
    \label{steering vector}
    \bm \alpha (\phi ^{(l)}) = \frac{1}{\sqrt{N_{BS}}} [1,  e^{-j2\pi \phi^{(l)}}, ... , e^{-j2\pi \phi^{(l)} (N_{BS}-1)}]^{T},
    \end{equation}	 
where $\phi ^{(l)}$ denotes the spatial direction of the $l$th path and is related to the physical angle $\theta^{(l)}$ by $\phi ^{(l)} = d/ \lambda \sin{\theta ^{(l)}}$, where $\lambda$ is the wavelength, and $d$ is the antenna spacing. For $d =\lambda/2$, we have $-\frac{1}{2} \leq \phi ^{(l)} \leq \frac{1}{2}$ and $-\frac{\pi}{2} \leq \theta ^{(l)} \leq \frac{\pi}{2}$.
The physical angle $\theta^{(l)}$ for the $l$th path is assumed to be uniformly distributed over $[-\pi/2, \pi/2 ]$. The NLoS path gain $\beta^{(l)}$ for $2 \leq l \leq N_p$ is assumed to follow the standard complex Gaussian distribution \bb{\cite{gao2016structured, Wei2021deep}}.
	
The spatial channel vector $\mathbf h$ can be transformed into the beamspace channel representation $\mathbf h_b$ using a discrete Fourier transform (DFT) as
	\begin{equation} 
	\label{h_b_u}
	\mathbf h_{b} = \mathbf{U} \mathbf h,
	\end{equation}    
where the DFT matrix $\mathbf U \in \mathbb{C}^{N \times N}$ is represented using a set of orthogonal array steering vectors as	$\mathbf{U}  = [\bm \alpha(\phi_1),  \bm \alpha(\phi_2),  ...  , \bm \alpha(\phi_{N})]^H$ \cite{brady2013beamspace}, where $\phi_i = \frac{1}{N} (i-\frac{N+1}{2})$ for $i=1,2,...,N$ is the virtual spatial direction predefined by the array having half-wavelength spaced antennas. 
The beamspace channel $\mathbf h_{b}$ is approximately sparse, which means that most of entries in $\mathbf h_{b}$ are nearly zeros. 
Since the lens antenna array can be designed to approximate an ideal DFT transmission \cite{brady2013beamspace}, \eqref{overline_r_m} can be written as 
\begin{equation}
\begin{aligned}
\mathbf r_{m} & = \mathbf W_m \mathbf U \mathbf h + \mathbf n_{m} \\
& = \mathbf W_m \mathbf h_{b} + \mathbf n_{m},
\end{aligned}
\end{equation}
where $\mathbf h_{b}$ is the beamspace channel. 

The pilot observations over $B$ time blocks are
\begin{equation} \label{stack_z}
\mathbf z = 
\begin{bmatrix}
\mathbf r_{1} \\
...\\
\mathbf r_{B}
\end{bmatrix} 
= 
\begin{bmatrix}
\mathbf W_1 \\
... \\
\mathbf{W}_B 
\end{bmatrix} 
\mathbf h_{b}
+ 
\begin{bmatrix}
\overline{\mathbf n}_{1} \\
...\\
\overline{\mathbf n}_{B}
\end{bmatrix}.
\end{equation}
By denoting $\mathbf W = [\mathbf W_1^T,..., \mathbf W_B^T]^T$ and $\mathbf n = [\mathbf n_{1}^T, ..., \mathbf n_{B}^T]^T$, the pilot observation in \eqref{stack_z} is compactly expressed as 
\begin{equation} \label{z_u}
\mathbf z = \mathbf \Phi \mathbf h_{b} + \mathbf n,
\end{equation}
where $\mathbf z \in \mathbb{C}^{M}$ and $M = BN_{RF}$ is the total number of measurements. The measurement matrix $\mathbf \Phi \in \mathbb{C}^{M \times N}$ consists of the analog combiners $\{\mathbf W_m \in \mathbb{C}^{N_{RF}\times N}, 1 \le m \le B\}$, which are enabled by one-bit phase shifters, and therefore, entries of $\mathbf \Phi$ can be modeled to follow a Bernoulli distribution and are randomly selected from $\{-\frac{1}{\sqrt{M}}, \frac{1}{\sqrt{M}} \}$.

The beamspace channel estimation in \eqref{z_u} is equivalent to the two real-valued equations: 
	\begin{equation}
	\label{real_sepa_imag2}
	\begin{aligned}
	\Re(\mathbf z) &= \mathbf \Phi \Re (\mathbf h_{b}) + \Re (\mathbf n) \\
	\Im(\mathbf z) &= \mathbf \Phi \Im (\mathbf h_{b}) + \Im (\mathbf n).
	\end{aligned}
	\end{equation}
We use the vector $\mathbf x$ to denote an arbitrary real-valued sample, i.e., $\mathbf x=\Re (\mathbf h_{b})$ or $\mathbf x=\Im (\mathbf h_{b})$, and \bb{use the vector $\mathbf y$ to denote a real-valued measurement vector, i.e., $\mathbf y=\Re (\mathbf z)$ or $\mathbf y=\Im (\mathbf z)$.} The sparse channel reconstruction problem is then expressed as
	\begin{equation} \label{linear_equation}
	\mathbf y = \mathbf \Phi \mathbf x + \mathbf n,
	\end{equation} 	
where $\mathbf y \in \mathbb{R}^{M}$, $\mathbf \Phi \in \mathbb{R}^{M \times N}$, $\mathbf x \in \mathbb{R}^{N}$, $\mathbf n \in  \mathbb{R}^{M}$, and $M \ll N$. 
	Since $\mathbf x$ is sparse, we can find it by solving
	\begin{equation} \label{SRopt}
	\begin{aligned}
	\min \limits_{\mathbf{x}} & \quad  \norm{\mathbf{y}-\mathbf{\Phi}\mathbf{x}}_2^2\\
	\text{s.t.} & \quad \norm{\mathbf{x}}_0  \le K.
	\end{aligned}
	\end{equation}

The $\ell_0$-norm constraint $\norm{\mathbf{x}}_0  \le K$ in \eqref{SRopt} results the problem NP-hard. 
A maximum a posteriori (MAP) estimator with a Laplacian prior, which is also known as the Lasso regression, i.e., $ \min \limits_{\mathbf{x}} 
	\norm{\mathbf{y}-\mathbf{\Phi}\mathbf{x}}_2^2 
	+ \lambda_1 \norm{\mathbf x}_1$, is a widely-used approximate to \eqref{SRopt}. 
Another approximate solution is the ridge regression, i.e., $
	\min \limits_{\mathbf{x}}  
	\frac{1}{2}\norm{\mathbf{y}-\mathbf{\Phi}\mathbf{x}}_2^2 
	+ \lambda_2 \norm{\mathbf x}_2^2$, which is a MAP estimator with a Gaussian prior but does not provide sparsity in general. \bb{The $\lambda_1$ and $\lambda_2$ denote regularization parameters on the $\ell_1$- and $\ell_2$-regularizers $\lambda_1 \norm{\mathbf x}_1$ and $\lambda_2 \norm{\mathbf x}_2^2$, respectively.}
A proposed solution here is a trimmed $\ell_2$-norm regularizer to remove the penalties on a few large-magnitude elements and promote sparse solutions. 

\bb{Since the large bandwidth is the main advantage of mmWave frequency band, it is reasonable to adopt the wideband channel model and system model for mmWave massive MIMO systems \cite{chen2019channel, chen2021hybrid}. Although this work focuses on narrowband block-fading channels of massive MIMO, the proposed algorithms are generic sparse reconstruction algorithms and can conveniently extend to the wideband massive MIMO channels which have sparse representations in a virtual angular-delay-Doppler domain.}

\subsection{Trimmed-Ridge Regression}	\label{trr} 
The top-$\ell_{K,2}$ norm permits reformulation of the sparse-reconstruction problem in \eqref{SRopt}. The top-$\ell_{K,2}$ norm \footnote{\bb{In our prior work \cite{wu2021nonconvex}, the top-(K, 1) norm was proposed to develop an iterative algorithm for sparse channel reconstruction. However, unfolding the iterative algorithm that contains top-(K, 1) norm led to unsatisfying performance.}} is the square sum of the $K$ elements having the largest magnitudes in the vector $\mathbf x$, i.e.,
	$\norm{\mathbf x}^2_{K,2} :=  x^2_{(1)} + x^2_{(2)} + \cdots + x^2_{(K)}$, 
where $x_{(i)}$ denotes the element whose magnitude is the $i$th-largest among the $N$ elements of the vector $\mathbf x$, i.e., $ | x_{(1)} | \ge| x_{(2)} | \ge \cdots \ge | x_{(K)} |$.
Since $\norm{\mathbf x}_2^2-\norm{\mathbf x}^2_{K,2}=0$ is equivalent to $\norm{\mathbf{x}}_0  \le K$, the sparse reconstruction optimization in \eqref{SRopt} can be rewritten as
	\begin{equation} \label{SRopt_topK2}
	\begin{aligned}
	\min \limits_{\mathbf{x}} & \quad  \norm{\mathbf{y}-\mathbf{\Phi}\mathbf{x}}_2^2\\
	\text{s.t.} & \quad \norm{\mathbf x}_2^2-\norm{\mathbf x}^2_{K,2}=0.
	\end{aligned}
	\end{equation}
Use of a Lagrange multiplier $\rho$ transforms the constrained optimization \eqref{SRopt_topK2} into an unconstrained optimization as
	\begin{equation} \label{SRopt_topK2_unconstrain}
	\min \limits_{\mathbf{x}} \quad 
	\frac{1}{2}\norm{\mathbf{y}-\mathbf{\Phi}\mathbf{x}}_2^2 
	+ \rho (\norm{\mathbf x}^2_2 - \norm{\mathbf x}^2_{K,2})
	:=F(\mathbf x).
	\end{equation}
The optimization in \eqref{SRopt_topK2_unconstrain} is referred to as the trimmed-ridge regression. 
This regularizer in \eqref{SRopt_topK2_unconstrain} helps achieve sparse and accurate solutions.


\section{Iterative Trimmed-Ridge Regression Algorithm}	

This section describes a trimmed-ridge regression algorithm that uses DC programming to derive an iterative solution. A discussion on determining the algorithm step size then follows. 

\subsection{Deriving an Iteration Solution Using DC Programming} 
\label{dcitrr}
The trimmed-ridge regression \eqref{SRopt_topK2_unconstrain} is a non-convex optimization.
The objective $F(\mathbf x)$ in \eqref{SRopt_topK2_unconstrain} can be equivalently written as  
 	\begin{equation} \label{obj_k2norm}
	F(\mathbf x)
	=
	 \frac{k}{2} \norm{\mathbf x}_2^2 -  \frac{k}{2} \norm{\mathbf x}_2^2 + \frac{1}{2}\norm{\mathbf y - \mathbf {\Phi x}}_2^2  + \rho (\norm {\mathbf x}_2^2 - \norm{\mathbf x}^2_{K,2}),
	\end{equation}
where $k \ge 0$ is a Lipschitz constant of the least squares objective $\frac{1}{2}\norm{\mathbf y - \mathbf {\Phi x}}_2^2$. 
	We decompose the objective function $F(\mathbf x)$ into convex functions $f(\mathbf x)$ and $g(\mathbf x)$ \footnote{The convexity of $h(\mathbf x)$ can be proved by showing it's Hessian matrix is semidefinite, i.e., $k \mathbf I - \mathbf \Phi^T \mathbf \Phi \succeq \mathbf 0$, for $k = \lambda_{max}(\mathbf \Phi^T \mathbf \Phi)$, where $\lambda_{max} (\cdot)$ denotes the largest eigenvalue of a matrix \cite{Peng2021thesis}.},
	\begin{equation}
	\label{obj_x_dc}
	\min_\mathbf x 
	\underbrace{ \frac{k}{2} \norm{\mathbf x}_2^2 + \rho \norm {\mathbf x}_2^2}_{f(\mathbf x)} - \underbrace{ \left( \frac{k}{2} \norm{\mathbf x}_2^2- \frac{1}{2}\norm{\mathbf y - \mathbf {\Phi x}}_2^2
	 + \rho \norm{\mathbf x}^2_{K,2} \right)}_{g(\mathbf x)}.
	\end{equation}
	
Next, the solution splits the positive and negative parts of $\mathbf x$ by letting $\mathbf u = (\mathbf x)_+$ and $\mathbf v = (-\mathbf x)_+$, where $(\mathbf x)_+$ represents the operation of $(x)_+ = \max \{0, x\}$ for each element $x$ in vector $\mathbf x$; $(-\mathbf x)_+$ represents the operation of $(-x)_+ = \max \{0, -x\}$ for each element $-x$ in vector $-\mathbf x$. Hence, $\mathbf u$ and $\mathbf v$ are nonnegative vectors.
Let $\mathbf{z} = [\mathbf{u}^{T}, \mathbf{v}^{T}]^{T}$, $\mathbf A = [\mathbf \Phi, -\mathbf \Phi]$, and express \eqref{obj_x_dc} as 
	\begin{equation} \label{obj_z_dc}
	\min_{\mathbf{z} \succeq \mathbf{0}} 
	\underbrace{ \frac{k}{2} \norm{\mathbf z}_2^2 + \rho \norm {\mathbf z}_2^2}_{f(\mathbf z)} - \underbrace{ \left( \frac{k}{2} \norm{\mathbf z}_2^2- \frac{1}{2}\norm{\mathbf y - \mathbf {A z}}_2^2
	 + \rho \norm{\mathbf z}^2_{K,2} \right)}_{g(\mathbf z)},
	\end{equation}
where ${\mathbf{z} \succeq \mathbf{0}}$ represents that $\mathbf{z}$ is a non-negative vector. 
DC programming solves \eqref{obj_z_dc} by repeating the following two steps until convergence: (a) compute the gradient (or subgradient) $ \partial g(\mathbf z^{(t)})$ for an arbitrary $t$th iteration; (b) solve the convex subproblem $\min \limits_{\mathbf{z}\succeq \mathbf{0}}  f(\mathbf z) - \mathbf z^T \partial g(\mathbf z^{(t)})$ to obtain $\mathbf z^{(t+1)}$. The two steps specifically are
	\begin{equation} 
	\label{dc_procedure}
	\begin{aligned}
	(\text{a}) & 
	\quad \partial g(\mathbf z^{(t)}) 
	=  k \mathbf{z}^{(t)} - (\mathbf y - \mathbf{A} \mathbf{z}^{(t)}) + 2 \rho \mathbf{\overline{z}}^{(t)}_{K} \\	
	(\text{b}) &
	\quad \mathbf z^{(t+1)} = \arg \min \limits_{\mathbf{z} \succeq \mathbf{0}} 
	  \big\{\frac{k}{2} \norm{\mathbf z}_2^2 + \rho \norm{\mathbf z}_2^2 - \mathbf z^T \partial g(\mathbf z^{(t)}) \big\}, 
	 \end{aligned}
	\end{equation}
where $\mathbf{\overline{z}}^{(t)}_{K}$ is the half of the gradient of top-$\ell_{K, 2}$ norm while noting that $\mathbf{\overline{z}}^{(t)}_{K}=z_{(1)} + z_{(2)} + \cdots + z_{(K)}$, where $z_{(i)}$ denotes the element whose magnitude is the $i$th-largest among the elements of the vector $\mathbf z$, i.e., $ z_{(1)}  \ge  z_{(2)}  \ge \cdots \ge z_{(K)} $.
The step (b) in \eqref{dc_procedure} can be expressed as 
	\begin{equation}
	\label{proj_prob_closed_z2}
	\begin{aligned}
	\min \limits_{\mathbf z \succeq \mathbf{0}} 
	& \quad
	\frac{k+2\rho}{2} \norm{\mathbf z - \frac{1}{k+2\rho}\partial g(\mathbf z^{(t)}) }_2^2.
	\end{aligned}
	\end{equation} 
A closed-form solution to \eqref{proj_prob_closed_z2} is the Euclidean projection of $\frac{1}{k+2\rho} \partial g(\mathbf z^{(t)})$ onto the nonnegative orthant, i.e.,
	\begin{equation} \label{optimal_z_closed}
	\mathbf z^* = \left( \frac{1}{k+2\rho} \partial g(\mathbf z^{(t)}) \right)_+,
	\end{equation} 
where $(\cdot)_+$ denotes the positive-retaining operation that sets the negative elements to zeros.
The DC programming procedure in \eqref{dc_procedure} reduces to
	\begin{equation} \label{dc_alg_closed}
	\begin{aligned}
	(\text{a}) &
	\quad  \partial g(\mathbf z^{(t)})  =  k \mathbf{z}^{(t)} - (\mathbf y - \mathbf{A} \mathbf{z}^{(t)}) + 2 \rho \mathbf{\overline{z}}^{(t)}_{K} \\	
	(\text{b}) &
	\quad \mathbf z^{(t+1)} 
	 = \left( \frac{1}{k+2\rho} \partial g(\mathbf z^{(t)}) \right)_+.
	 \end{aligned}
	\end{equation} 
Thus, \eqref{dc_alg_closed} is an iterative solution to the trimmed-ridge regression in \eqref{SRopt_topK2_unconstrain}.
Algorithm 1 summarizes the iterations in \eqref{dc_alg_closed}. Since the iterations are derived under a DC programming framework, Algorithm 1 shares the same global convergence property as the general DC programming \cite{Thi2018DC}.
    
\begin{algorithm}
	\caption{Iterative trimmed-ridge regression (ITRR)} 
	\hspace*{\algorithmicindent} \textbf{Input:} measurements $\mathbf y$, matrix $\mathbf A$ and a small number $\epsilon$ \\
    \hspace*{\algorithmicindent} \textbf{Output:} reconstructed $\hat{\mathbf x}$ \\
    	\hspace*{\algorithmicindent} \textbf{Initialization:}  $\mathbf u^{(0)}$, $\mathbf v^{(0)}$, $\mathbf z^{(0)} \leftarrow [(\mathbf u^{(0)})^T,  (\mathbf v^{(0)})^T]^T$
	\begin{algorithmic}[1]
		\For {$t=0,1,\ldots$}
			\State $\nabla F(\mathbf z^{(t)}) \leftarrow \mathbf{A}^T (\mathbf y - \mathbf{A} \mathbf{z}^{(t)})+  2 \rho (\mathbf z^{(t)}-\mathbf{\overline{z}}^{(t)}_{K})$
			\State $\mathbf{z}^{(t+1)} \leftarrow \left(\mathbf{z}^{(t)} - \frac{1}{k+2\rho} \nabla F(\mathbf{z}^{(t)})\right)_+$ 
			\State Check termination condition $\norm{\mathbf z^{(t+1)}-\mathbf z^{(t)}}_2 \le \epsilon$, return to Step 1 if not satisfied; otherwise, terminate with $\mathbf z^{(t+1)} =  [(\mathbf u^{(t+1)})^T,  (\mathbf v^{(t+1)})^T]^T$, and return $\hat{\mathbf x} = \mathbf u^{(t+1)} - \mathbf v^{(t+1)}$.
		\EndFor
	\end{algorithmic} 
\end{algorithm}


\subsection{Relaxing Step Sizes}
\label{BBstepsize}
Step 2 and Step 3 in Algorithm 1 can be combined to  write the update for $\mathbf z^{(t+1)}$ as
\begin{equation}
	\label{interpret_t_update1}
	\begin{aligned}
	\mathbf z^{(t+1)}  
	& = \left( \mathbf z^{(t)} - \frac{1}{k+2\rho}\underbrace{ \left(\mathbf{A}^T (\mathbf y - \mathbf{A} \mathbf{z}^{(t)})+ 2 \rho (\mathbf z^{(t)}-  \mathbf{\overline{z}}^{(t)}_{K}) \right )}_{\nabla F(\mathbf z^{(t)})} \right )_+.
	\end{aligned}	
\end{equation}
We observe that \eqref{interpret_t_update1} is a gradient-projection descent update with a step size of $1/(k+2\rho)$ by defining $F(\mathbf z^{(t)}) = \frac{1}{2}\norm{\mathbf{y}-\mathbf{A}\mathbf{z}}_2^2 + \rho (\norm{\mathbf z}^2_2 - \norm{\mathbf z}^2_{K,2})$. 
The fixed step size $1/(k+2\rho)$, which is tied to the Lipschitz constant $k$, can lead to slow convergence. 
This motivates use of an accelerated step size strategy \cite{Parikh2014proximal}. The monotonic Barzilai-Borwein (BB) step size is adopted, where the $t$th-step update becomes
    \begin{equation} \label{mono_GPD}
    \begin{aligned}
    \mathbf{w}^{(t+1)} &= \left(\mathbf{z}^{(t)} - \alpha^{(t)} \nabla F(\mathbf{z}^{(t)})\right)_+, \\
     \mathbf{z}^{(t+1)} &= \mathbf{z}^{(t)} + \beta^{(t)} ( \mathbf{w}^{(t+1)}  -  \mathbf{z}^{(t)} ).
     \end{aligned}
    \end{equation}
Here $\alpha^{(t)} > 0$ is the step size for the $t$th update and $\beta^{(t)}\in (0, 1]$ is the extrapolation parameter that prevents the step size from being too large to ensure the objective value descends monotonically.
The step size $\alpha^{(t)}$ is explicitly computed by 
	\begin{equation}
    \alpha^{t} = \frac{\norm{\mathbf z^{t}- \mathbf z^{t-1}}^2}{(\mathbf z^{t} - \mathbf z^{t-1})^T \left( F(\mathbf z^t) - F(\mathbf z^{t-1}) \right)}.
	 \end{equation}
The extrapolation parameter $\beta^t$ is often empirically set as $t/(t+3)$ \cite{Parikh2014proximal}. 
Due to the nonconvex property of $F(\mathbf z^{(t)})$, the iteration described in \eqref{mono_GPD} lacks a theoretical guarantee of convergence to the optimal solution. 
However, Section \ref{SimSec}'s simulations demonstrate that the iterative algorithm with the BB step sizes achieves a consistently satisfactory level of reconstruction accuracy.

\subsection{ITRR Performance on Reconstructing a Sparse Sample}
\label{e_itrr_Sam}
This section shows the convergence property and the reconstruction accuracy of the ITRR algorithm when reconstructing a sparse sample. 
Simulations artificially produced a sparse sample, $\mathbf h_s$ with $16$ nonzero entries from an arbitrary beamspace channel vector, by setting all other elements to zero except for the $16$ largest-magnitude entries. The measurements $\mathbf \Phi \mathbf h_s$ were the input to obtain the reconstruction $\hat{\mathbf h}_b$. Both the ITRR using a fixed step size and the ITRR using the BB step size (ITRR-BB) were tested. The Lasso and ridge regressions were performed to reconstruct the same sparse sample using a similar iterative algorithm, i.e., the gradient-projection descent using the BB step size (PGD-BB).
\bb{Only the results of the real part of $\mathbf h_s$ will be demonstrated. The results for the imaginary part are similar, because we treat the real part the imaginary part of a channel as two independent real-valued vectors as in \eqref{real_sepa_imag2} and \eqref{linear_equation}.}

\begin{figure*}[!t]
\centering
{\includegraphics[scale=0.33]{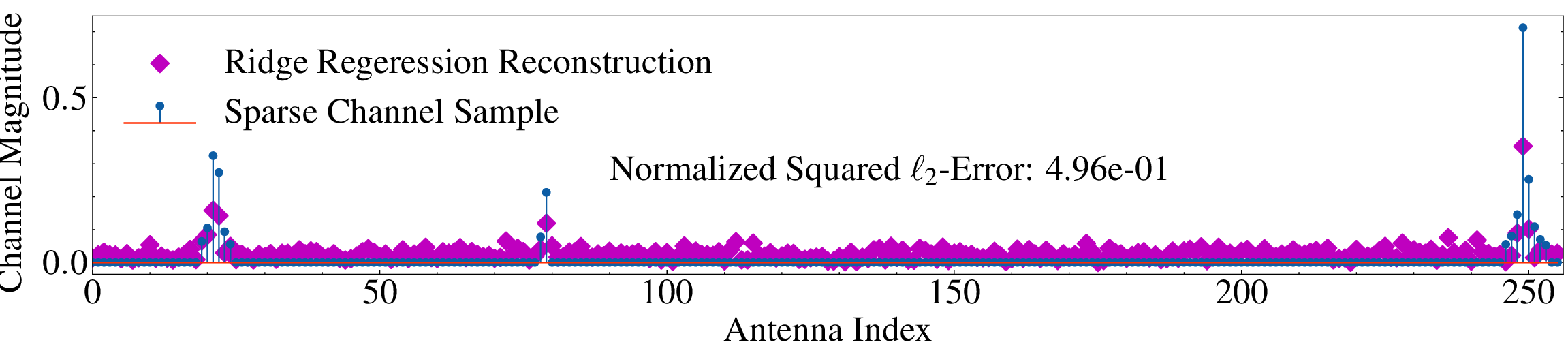}}\\
{\includegraphics[scale=0.33]{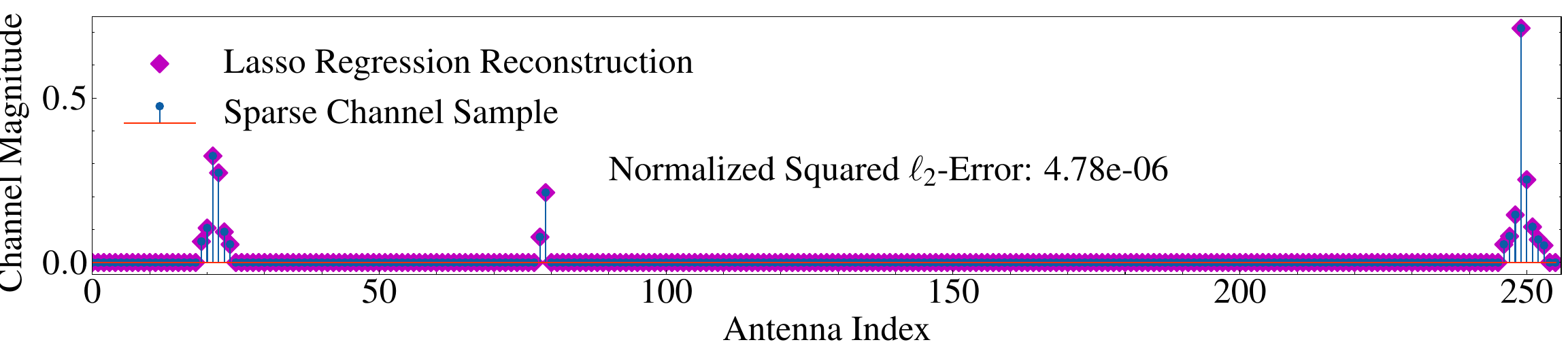}}\\
{\includegraphics[scale=0.33]{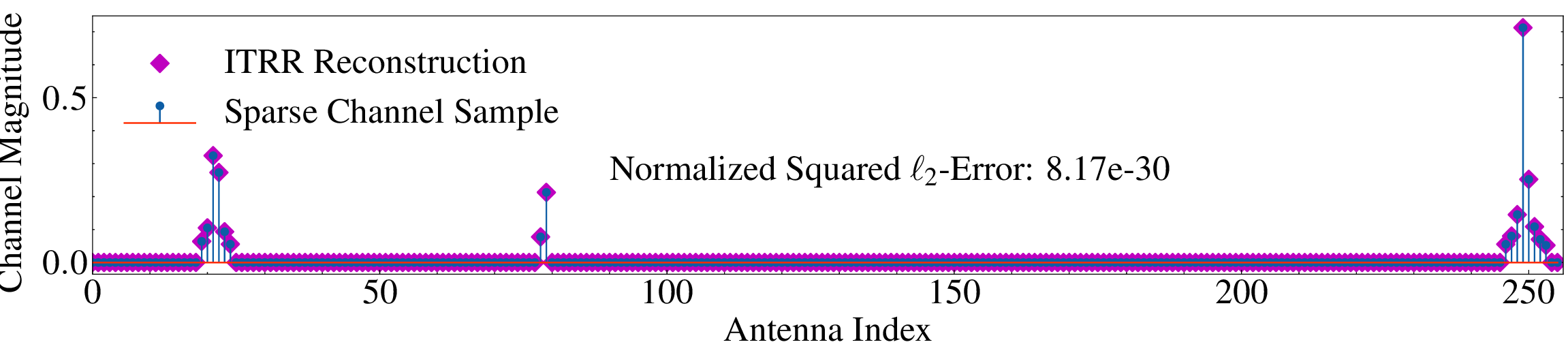}} \\
\caption{Channel magnitudes of a sparse sample $\mathbf h_s$ and the reconstruction $\hat{\mathbf h}_s$ by the ridge regression, Lasso regression and the proposed ITRR algorithm, wherein the normalized squared $\ell_2$-error is computed as $\norm{\Re(\mathbf{h}_s)-\Re(\hat{\mathbf{h}}_s)}_2^2/\norm{\Re(\mathbf h_s)}_2^2$.}
\label{sparse_recon_sample}
\end{figure*}

Figure \ref{sparse_recon_sample} shows the reconstructions against the true sparse sample.
The ridge regression produces a noisy and inaccurate reconstruction having the normalized $\ell_2$-error in the order of magnitude of $-1$.
The Lasso regression produces a sparse reconstruction having a normalized $\ell_2$-error in the order of magnitude of $-6$. The proposed ITRR algorithm produces an exact sparse reconstruction in the order of magnitude of $-30$. Both the ITRR and the ITRR-BB algorithms achieved similar reconstruction results in our numerical experiments.
Optimization theory explains the difference in reconstruction accuracy: The $\ell_1$-regularizer encourages sparsity but the same penalty on all element degrades accuracy. 
\bb{On the contrary, $\ell_2$-regularizer discourages sparsity. Since $\ell_2$-regularizer gradient is $\nabla \norm{\mathbf x}_2^2 = 2\mathbf x$, the gradients of small absolute-value elements are small. Therefore, $\ell_2$-regularizer yields diminishing gradients when elements values reduce to nearly zero, thus, the algorithm is more likely to converge before achieving a sparse solution.} The proposed trimmed-ridge regression removes the penalties on large-magnitude elements to promote both the sparsity and accuracy.

\begin{figure}[t]
\centering
\normalsize
\includegraphics[width=3.2in]{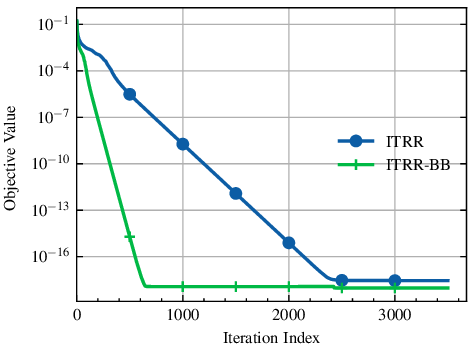}
\caption{Objective evolution when reconstructing the real part of the sparse sample $\Re(\mathbf h_s)$. }
\label{evo1}
\end{figure}

\begin{figure}[t]
\centering
\normalsize
\includegraphics[width=3.2in]{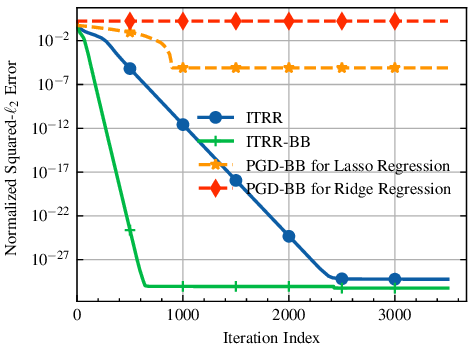}
\caption{Normalized reconstruction error evolution (i.e., $\norm{\hat{\Re(\mathbf h_s)}-\Re(\mathbf h_s)}_2^2 / \norm{\Re(\mathbf h_s)}_2^2 )$) when reconstructing the real part of the sparse sample}
\label{evo2}
\end{figure}

\begin{figure}[t]
\centering
\normalsize
\includegraphics[width=3.2in]{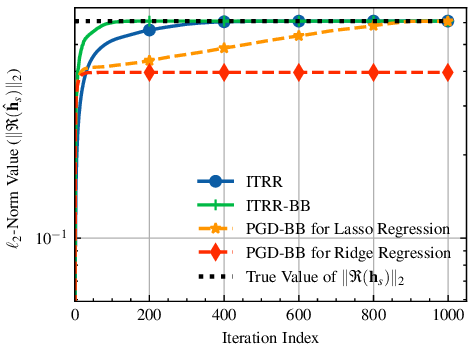}
\caption{The $\ell_2$-norm value evolution when reconstructing the real part of the sparse sample}
\label{evo3}
\end{figure}

Figures \ref{evo1}--\ref{evo3} demonstrate the convergence properties of the iterative algorithms when constructing the real part of a sparse sample.
Fig. \ref{evo1} shows the objective values versus iterations for ITRR and ITRR-BB algorithms.
Fig. \ref{evo2} shows the normalized $\ell_2$-error of reconstructions over iterations. The ITRR and ITRR-BB reached lower reconstruction errors on the orders of magnitude of $-30$ and $-31$, while the Lasso and ridge regression algorithms achieved errors on the orders of magnitude of $-6$ and $-1$.
The objective evolution in Fig. \ref{evo1} and the reconstruction error evolution in Fig. \ref{evo2} together suggest that using the BB steps size in an ITRR algorithm can accelerate the convergence without degrading the reconstruction accuracy.
Fig. \ref{evo3} shows the $\ell_2$-norm values of the reconstruction vectors versus iterations, where the true $\ell_2$-norm value is also plotted for comparison. 
The reconstructions by the ITRR-BB, the PGD-BB for Lasso regression, and the ITRR in turn achieved the optimal $\ell_2$-norm, whereas the PGD-BB for ridge regression left a constant gap between the reconstruction $\ell_2$-norm and the optimal $\ell_2$-norm. 
The ridge regression has an $\ell_2$-regularizer that discourages sparsity and thus leads to an inaccurate reconstruction, while the proposed ITRR and ITRR-BB algorithm achieves an accurate solution.
These results demonstrate the advantage of using the proposed top-$\ell_{K,2}$ to remove the penalties on $K$ large magnitude elements from the conventional $\ell_2$-regularizer.

\section{Learned Trimmed-Ridge Regression Algorithm}
While the ITRR algorithm offers the potential for direct beamspace channel estimation, its practical implementation can pose significant challenges.
First, the computational complexity of ITRR increases rapidly with the problem dimension, i.e., the number of BS antennas of a massive MIMO system.
Second, the regularization parameter $\rho$ and the top-$K$ parameter ideally should be tuned specifically for each channel realization, because using a fixed parameter setting can be suboptimal and consequently result in a degradation of the reconstruction performance. 
These two properties, which are not exclusive to ITRR but are commonly found in iterative and regularized regression algorithms, can result in slow convergence, making them unsuitable for real-time applications.
A mmWave massive MIMO system has limited channel coherence time and thus requires low latency for channel-estimation algorithms.
A deep-learning-based method is a promising alternative for practical online implementation due to its non-iterative and the inherent parallel computation ability. 
This motivates development of a model-based deep-learning approach that emulates the proposed ITRR algorithm, primarily because the ITRR-iteration qualities make it easy to comprehend, exhibit a straightforward computational structure, and demonstrate competitive reconstruction accuracy. The following sections first present the basic deep-network structure and then propose important configurations to reduce the computational complexity and a model-ensemble method to improve the reconstruction accuracy.

\subsection{Unfolding Trimmed-Ridge Regression (UTRR) Structure}
\label{utrr}

\begin{figure*}[!t]
\centering
\includegraphics[scale=0.45]{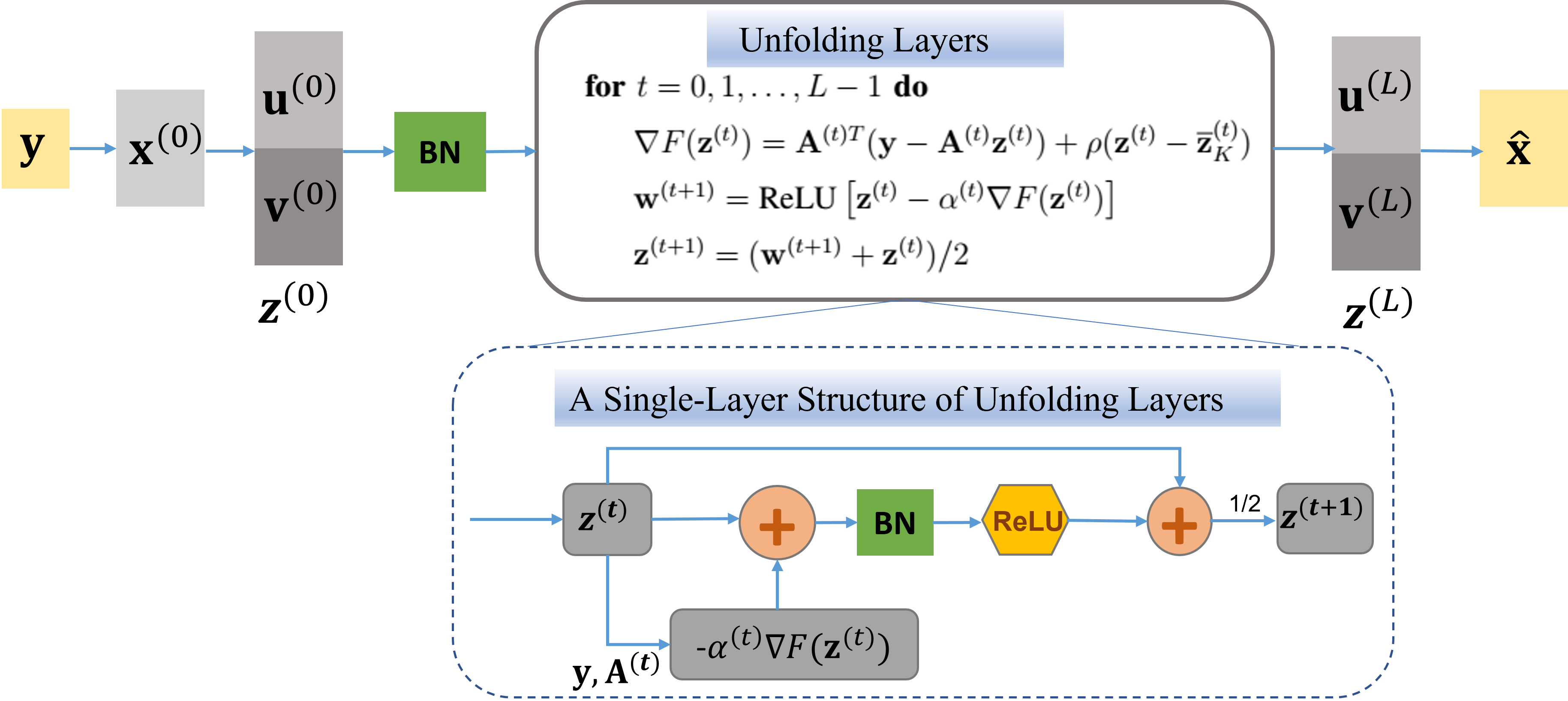}
\captionsetup{justification=centering}
\caption{The model structure of UTRR. BN represents a layer of Batch Normalization.}
\label{gpsa_structure}
\end{figure*}

As in Fig. \ref{gpsa_structure}, the UTRR framework consists of a \mbox{single-layer} encoder from $\mathbf x$ to $\mathbf y$ and a \mbox{multi-layer} decoder from $\mathbf y$ to $\hat{\mathbf x}$. \bb{The encoder $\mathbf{y} = \mathbf{\Phi} \mathbf{x}$ performs a linear dimension reduction. The input sparse vector $\mathbf{x}$ is projected onto the subspace spanned by the columns of the measurement matrix $\mathbf{\Phi}$ to obtain the compressed measurement vector $\mathbf{y}$.} The decoder plays the role of the sparse reconstruction from $\mathbf{y}$ to $\hat{\mathbf{x}}$ in a feed-forward manner. The decoder is constructed to mimic the ITRR to obtain the reconstructed $\hat{\mathbf{x}}$. We can interpret the decoder by three functional parts, including the initializing layer, the unfolding layers and the output layer.
	
\emph{Initializing layer ($\mathbf y$ to $\mathbf z^{(0)}$):} The first two layers can be understood as the initialization of ITRR, which originates from the input, i.e., the measurements $\mathbf{y}$ and obtain the initialized positive vector $\mathbf{z}^{(0)}$. We first initialize $\mathbf{x}^{(0)}$ to be $ \mathbf{x}^{(0)}= \mathbf{\Phi}^T \mathbf y$. Then, we prepare an equivalent positive initialization vector $\mathbf{z}^{(0)}$ by separating and concatenating the positive and negative part of $\mathbf{x}^{(0)}$, that is $\mathbf{z}^{(0)}=[(\mathbf{u}^{(0)})^T, (\mathbf{v}^{(0)})^T]^T$, where $\mathbf{u}^{(0)} = \text{ReLU}(\mathbf{x}^{(0)})$ and $\mathbf{v}^{(0)} = \text{ReLU}(-\mathbf{x}^{(0)})$.
	
	\emph{Unfolding layers ($\mathbf{z}^{(1)}$ to $\mathbf{z}^{(L)}$):} 
	In unfolding layers, the iterations of the ITRR algorithm are stacked into multiple layers of the deep network by treating each iterative update step as a deep-network layer.	
	The unfolding layers perform a forward update from $\mathbf{z}^{(1)}$ to $\mathbf{z}^{(L)}$, which consists of $L$ \mbox{structural-identical} blocks. Each block is a one-step iteration of the ITRR algorithm using the BB step-size structure. However, the proposed algorithm sets the step size $\alpha^{(t)}$ as a trainable parameter and sets the extrapolation parameter to be $\beta^{(t)} =  1/2$. 
	This setting is equivalent to designing a residual learning unit \cite{he2016deep} within each unfolding layer.
	The trainable weights of the $t$th unfolding layer are in $\mathbf A^{(t)} \in \mathbb{R} ^{M \times 2N}$, which is initialized as $[\mathbf{\Phi}$, $-\mathbf{\Phi}]$.
	The ReLU activation function in the unfolding layers is notably equivalent to the projection operations $(\cdot)_+$ in the ITRR updates.

\emph{Output layer ($\mathbf z^{(L)}$ to $\hat{\mathbf x}$):} The last layer is the output layer to obtain the reconstructed vector $\hat{\mathbf{x}}$. Recalling that $\mathbf{z}^{(L)} = [z^{(L)}_1, ..., z^{(L)}_{2N}]$; therefore, we have
	$
	\hat{\mathbf{x}}=\mathbf{u}^{(L)}- \mathbf{v}^{(L)}
	$, where $\mathbf{u}^{(L)} = [z^{(L)}_1, ..., z^{(L)}_{N}]$ and $\mathbf{v}^{(L)} = [z^{(L)}_{N+1}, ..., z^{(L)}_{2N}]$.

\subsection{Reducing Computational Complexity}
\label{sec_rcc}

The basic UTRR model structure in Fig. \ref{gpsa_structure} helps specify the configurations in the unfolding layers to reduce computational complexity and improve model prediction accuracy.
\bb{A UTRR model has multiple identical unfolding layers from $\mathbf z^{(1)}$ to $\mathbf z^{(L)}$, and each layer contains a $\mathbf{\overline{z}}^{(t)}_{K}$ term which is the gradient of the top-$\ell_{K,2}$ norm $\norm{\mathbf z}^2_{K,2}$.} 
The $\mathbf{\overline{z}}^{(t)}_{K}$ calculation must sort all elements of the vector $\mathbf z^{(t)}$. 
If such sorting operation is performed on every training sample within each layer, it causes a high computational complexity. 
To reduce this computational complexity, the top-$K$ parameter is set to zero for all the intermediate layers from $\mathbf z^{(1)}$ to $\mathbf z^{(L-1)}$ (except for the last layer $\mathbf z^{(L)}$). Thus, the unfolding iterations for the intermediate layers simplify to 
	\begin{equation} \label{rcc}
	\begin{aligned}
	\mathbf w^{(t+1)}  
	& =  \text{ReLU} \big( \mathbf z^{(t)} - \alpha^{(t)}  \left(\mathbf{A}^{(t)T} (\mathbf y - \mathbf{A}^{(t)} \mathbf{z}^{(t)})+ \rho \mathbf z^{(t)} \right)\big) \\
	\mathbf z^{(t+1)}  & =  (\mathbf w^{(t+1)} + \mathbf z^{(t)})/2
	\end{aligned}	
	\end{equation} 
for $0 \le t \le L-2$.
Eq. \eqref{rcc} shows the $\mathbf{\overline{z}}^{(t)}_{K}$ term has been directly removed for the intermediate layers. Thus the computation of $\mathbf w^{(t+1)}$ for $t \in \{0, 1, ..., L-2\}$ can be simply interpreted as unfolding the projected gradient descent on traditional ridge regression. 
This configuration has negligible impact on the model-prediction performance and can reduce the computational complexity significantly (as shown in Section \ref{e_rcc}).

\vspace{-2mm}
\subsection{Training Procedure}
The goal of training a UTRR model is to produce an output vector $\hat{\mathbf x}$ from the noisy measurement vector $\mathbf y$ such that $\hat{\mathbf x} \approx \mathbf x$, where $\mathbf x$ represents the true channel vector. 
To this end, we should use a supervised learning training technique and prepare a dataset $\mathcal{D} = \{(\mathbf x_1, \mathbf y_1), ..., (\mathbf x_n, \mathbf y_n) \}$, where $\mathcal{D}_y =\{\mathbf y_1, ..., \mathbf y_n \}$ include the noisy measurements and act as the training inputs; $\mathcal{D}_x=\{\mathbf x_1, ..., \mathbf x_n \}$ are the original channel samples and act as training labels. 
Given the dataset, the training process minimizes a loss function by optimizing the network weights (trainable parameters) using a backpropagation method. The loss function and trainable parameters are defined as follows:	

 \emph{Loss function:} The loss function is defined by the mean square error of the reconstruction as  
	 \begin{equation}
	 \label{loss_function_cha5} 
	 \mathcal{L}(\mathbf{x}, \hat{\mathbf{x}}) = \mathbb{E} [\norm{\mathbf x -\hat{\mathbf x}}_2^2].
	 \end{equation}

	\emph{Trainable parameters:} \bb{The trainable parameters are $\{\mathbf A^{(t)}, \rho^{(t)}, \alpha^{(t)} \}$ for $t \in \{0, ..., L-1\}$. The number of trainable parameters for a UTRR model would be $L(2MN+2)$.} Here, $\mathbf  A^{(t)}$ is initialized to $[\mathbf \Phi, -\mathbf \Phi]$, where $\mathbf \Phi$ is the measurement matrix. 
\bb{In this work, we simply initialize the parameter $\rho^{(t)}$ to $1.0$ and initialize the parameter $\alpha^{(t)}$ to $0.1$ for all layers and for all models. Note that those initialization values for $\rho^{(t)}$ and $\alpha^{(t)}$ may not be the optimal. Tunning techniques for those hyperparameters to achieve their optimality are out of the scope of this paper. Practical cross validation strategies can help to choose proper the initialization values for different dataset.} 
	Given the training dataset, the trainable parameters use backpropagation to minimize the loss function in \eqref{loss_function_cha5}.

Algorithm 2 summarizes a UTRR model training procedure. The Step 3 to Step 11 perform forward computations of the UTRR model, and Step 12 performs the backward propagation to optimize the network trainable parameters, which can be performed automatically by deep learning platforms (e.g., Tensorflow and Pytorch).

\begin{algorithm}
	\caption{UTRR Model Training Procedure} 
	\hspace*{\algorithmicindent} \textbf{Input:} Dataset $\mathcal D = \{(\mathbf x_1, \mathbf y_1), ..., (\mathbf x_n, \mathbf y_n) \}$ \\
    \hspace*{\algorithmicindent} \textbf{Output:} Optimized $\{\mathbf A^{(t)}, \rho^{(t)}, \alpha^{(t)} \}$ for $t \in \{0, ..., L-1\}$ \\
    	\hspace*{\algorithmicindent} \textbf{Initialize:}  Set top-$K$ parameter, initilize trainable parameters
	\begin{algorithmic}[1]
	\For {$\text{epoch} = 1, 2, \ldots$}
	\For {$\text{batch} = 1, 2, \ldots$}	
	\State $\mathbf{x}^{(0)} \leftarrow \mathbf{\Phi}^T \mathbf y$ 
	\State $\mathbf{z}^{(0)} \leftarrow [(\mathbf{x}^{(0)})_+^T, (-\mathbf{x}^{(0)})_+^T]^T$
		\For {$t=0,1,\ldots, L-1$}
			\State $\nabla F(\mathbf{z}^{(t)}) = \mathbf{A}^{(t)T} (\mathbf y - \mathbf{A}^{(t)} \mathbf{z}^{(t)})+ \rho (\mathbf z^{(t)} - \mathbf{\overline{z}}^{(t)}_{K})$
			\State $\mathbf w^{(t+1)} = \text{ReLU} \left[ \mathbf z^{(t)} - \alpha^{(t)} \nabla F(\mathbf{z}^{(t)}) \right] $ 
			\State $\mathbf z^{(t+1)}  =  (\mathbf w^{(t+1)} + \mathbf z^{(t)})/2$ 
			\EndFor
		\State For $\mathbf{z}^{(L)} = [z^{(L)}_1, ..., z^{(L)}_{2N}]$, set $\mathbf{u}^{(L)} = [z^{(L)}_1, ..., z^{(L)}_{N}],\mathbf{v}^{(L)} = [z^{(L)}_{N+1}, ..., z^{(L)}_{2N}]$	
		\State $\hat{\mathbf{x}} \leftarrow \mathbf{u}^{(L)}- \mathbf{v}^{(L)}$ 
		\State Minimize $\mathcal{L}$ in \eqref{loss_function_cha5} and update parameters $\{\mathbf A^{(t)}, \rho^{(t)}, \alpha^{(t)} \}$ for $t \in \{0, ..., L-1\}$  
	\EndFor
	\EndFor
	\end{algorithmic} 
\end{algorithm}

\subsection{Model-Averaging Ensemble}
\label{ensemble}
The previous sections have explained the model structure and training procedure for a UTRR model. This section proposes averaging an ensemble of multiple UTRR models to improve further the reconstruction accuracy. 
Model-averaging ensemble was originally proposed to train multiple models over smaller-size subsets of data, and then combine the multiple prediction results as the final prediction result \cite{Sollich1995learning}. 
Nowadays, machine-learning engineers use model-averaging ensemble to reduce model uncertainty (to reduce prediction variance without increasing prediction bias) \cite{ensemble1999liu}, because a model ensemble often performs better than any individual model (intuitively, errors from various models are averaged). 
The simplest ensemble method averages multiple models' prediction results; alternatively, a weighted-average method can weigh models unequally according to their performance. The model weights can be optimized by a basic cross-validation method or by advanced algorithms. For example, a ``meta-network" can be trained to treat each individual model as a ``neuron" and serve the synaptic weights of the trained network as the optimized ensemble weight \cite{improving2022jose}.  
This work explores the simplest strategy, i.e., averaging the sum of each model reconstruction result by using constant and equal weights. 
Extending our work to find the optimal model-averaging-ensemble strategy remains an interesting research direction. 

\begin{figure}[!t]
\centering
\includegraphics[scale=0.45]{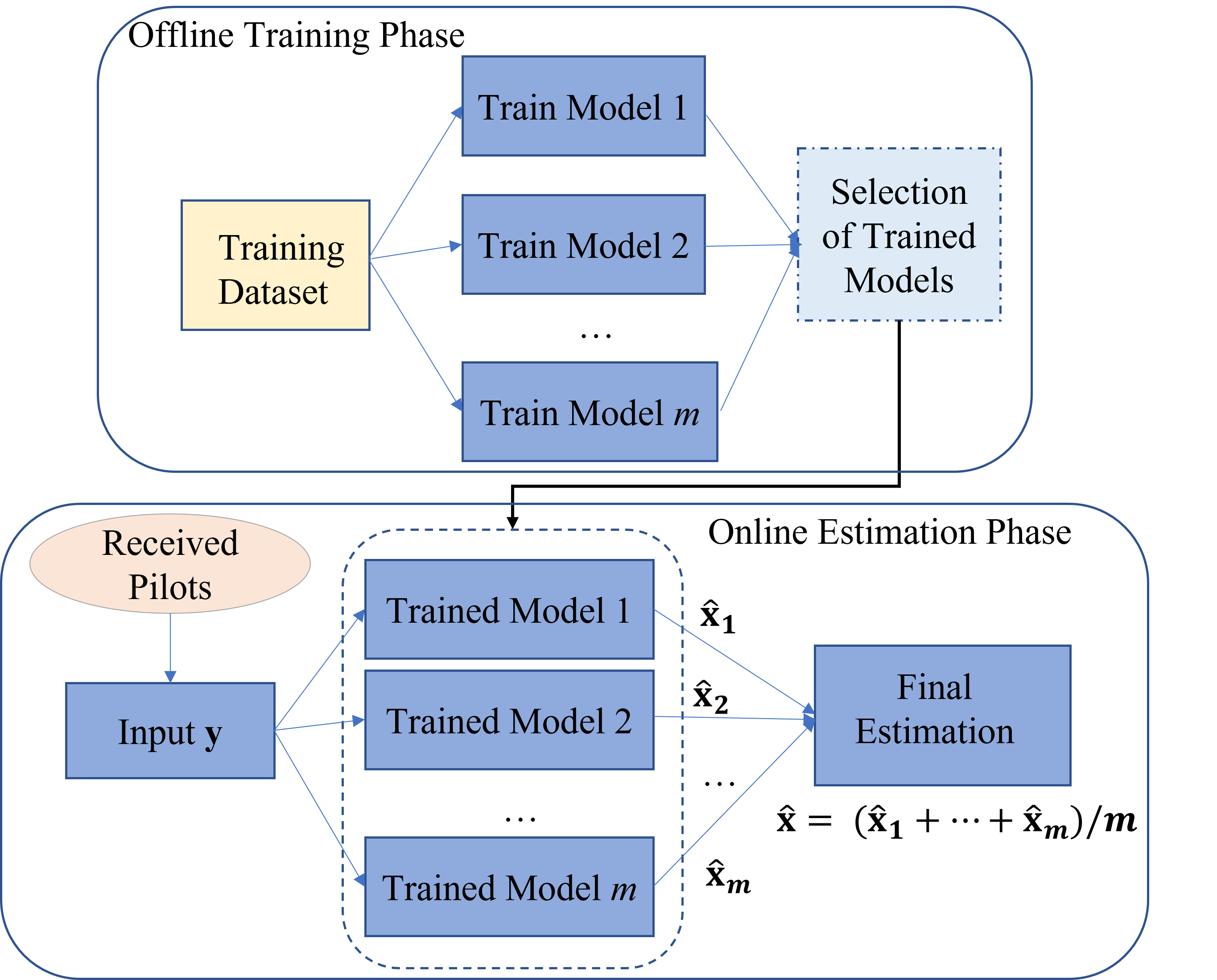}
\captionsetup{justification=centering}
\caption{Pipeline of using UTRR model-averaging ensemble for beamspace channel estimation}
\label{pipeline}
\end{figure}

Since multiple diverse models are required to develop a model-averaging ensemble, we should first prepare several models that produce reconstructions having small bias and large variance. 
The UTRR is readily available to develop diverse models because the $\mathbf{\overline{z}}^{(t)}_{K}$ term in the last unfolding layer (i.e., $t=L-1$) can be assigned with different values on the top-$K$ parameter.
The top-$K$ parameter implies the assumed sparsity level on the trimmed-ridge regression problem.
Then, multiple UTRR models with different top-$K$ parameters can be assembled for final reconstruction.
Recall that the top-$K$ parameters are zeroed for the intermediate unfolding layers to avoid computing the top-$\ell_{K,2}$ norm derivative term, whereas the term $\mathbf{\overline{z}}^{(t)}_{K}$ in the last unfolding layer (i.e., from $\mathbf z^{(L-1)}$ to $\mathbf z^{(L)}$) plays an important role in a model-ensemble approach. 
\bb{In} summary, a model-averaging ensemble includes three steps: 1) \bb{develop several UTRR models that have different values of the top-$K$ parameters}; 2) train each of the UTRR models using the same set of training data; 3) reconstruct the unseen beamspace channel by using each model and average their reconstructions. 
Fig. \ref{pipeline} shows the pipeline of using the UTRR model-averaging ensemble for beamspace channel estimation. 
The proposed UTRR is first trained offline to acquire the optimized model, and the optimized models are ensembled and implemented online to estimate the beamspace channels in real time.
The third step should perform parallel computations on all the model reconstructions to avoid extra latency in a practical implementation.

\section{Numerical Results}
\label{SimSec}
This section reports the main numerical results relevant to the proposed algorithms. Simulation setups and performance metrics will be first discussed. Then, performance of the ITRR algorithm and the UTRR models will be evaluated and compared with other methods. Also, this section evaluates the UTRR performance enhancement by the proposed computational-complexity reducing strategy and the model-averaging ensemble.

\subsection{Simulation Setup}
\label{SimSetup}

Simulations consider a mmWave massive MIMO system that has $U=16$ single-antenna users and a BS equipped with $N = 256$ antennas and $N_{RF}=16$ RF chains. 
To estimate the channels for all users, each user repeatedly transmits a pilot sequence having the length of $T=16$ over $B=8$ blocks. Thus, the BS receives $M=128$ measurements for each user, as described in \eqref{z_u}. Note that $M=128$ is a default setting in our simulation unless explicitly denoting different values of $M$.  
Following these system parameters, we first use the Saleh-Valenzuela channel model in \eqref{h_u} to generate the spatial channel samples $\{\mathbf h_1, ..., \mathbf h_n \}$, and then use a DFT in \eqref{h_b_u} to transform the spatial channel samples into the beamspace channel samples $\{\mathbf h_{b,1}, ..., \mathbf h_{b,n} \}$. 
Each channel sample is normalized to eliminate the effect of large-scale fading. 
The measurements $\{\mathbf z_1, ..., \mathbf z_n\}$ are produced according to \eqref{z_u} by $\mathbf z_i = \mathbf \Phi \mathbf h_{b,i} + \mathbf n_i$ for $1 \le i \le n$,  where $\mathbf \Phi \in \mathbb{R}^{M \times N}$ is a random matrix whose entries follow a Bernoulli distribution and are randomly selected from $\{-\frac{1}{\sqrt{M}}, \frac{1}{\sqrt{M}}\}$, and $\mathbf n_i$ is a Gaussian noise vector with a zero mean and a variance that is associated with the channel estimation signal-to-noise ratio (SNR), which is defined by $SNR=\mathbb{E}\left[\norm{\mathbf{\Phi} \mathbf h_b}^2 / \norm{\mathbf n}^2 \right]$. 
The training dataset, validation dataset and testing dataset contain $80,000$, $2,000$, and $2,000$ beamspace channels and measurements.

\bb{Experiments trained upon several UTRR models using different top-K parameters. The models were trained and evaluated individually, and then the trained models are used with model-averaging ensemble. In practice, if the sparsity level is known or estimatable, it is suggested to select different top-K parameters around the known/estimated sparsity level. In our simulations, we do not know the sparsity level of the beamspace channels since all the channels are approximately sparse instead of exact sparse. We randomly selected four top-K parameters $K \in \{0, 16, 64, 256\}$ for four UTRR models. Note that those model selections may not be the optimal choices. All UTRR models trained in this paper have 30 unfolding layers with the same training hyperparameters as follows. The diminishing leaning rates are adopted in the training process, and the learning rates are set to be $0.005, 0.001, 0.0005, 0.0002$, and $0.0001$. The maximum training epoch is set to be $300$. Early stopping is adopted to avoid overfitting, and the tolerate epoch number is set to 10, which means training stops if the validation error has not decreased for 10 epoches. The batch size is set to be $128$.}

Analyses use the normalized mean square error (NMSE) to evaluate channel estimation accuracy and compute the downlink sum rate to evaluate the influence of imperfect channel estimation on the communication-link capacity. The channel estimation NMSE is defined by $\mathbb{E}\left[\frac{(\norm{\mathbf h_{b} -\hat{\mathbf h}_{b}}_2^2)}{\mathbf h_{b}}\right]$, where $\mathbf h_{b}$ denotes the true beamspace channel and $\hat{\mathbf h}_{b}$ denotes the estimated beamspace channel. 
The downlink sum rate is $\mathbb{E}\left[ \sum\limits_{u=1}^{U} \log_2(1+\frac{\norm{\mathbf h_{b,u}^H \mathbf f_{u}}^2 }{\sum\limits_{i \neq u} \norm{\mathbf h_{b,u}^H \mathbf f_i}^2 + 1/{SNR}_{dl}}) \right]$, where $\mathbf f_{u}$ denotes the precoder for the $u$th user, and $SNR_{dl}$ denotes the downlink transmit SNR. \bb{For the downlink transmission that uses zero-forcing precoding, the transmission model can be represented as $y_u = \sqrt{P_u}  \mathbf h_u^T  \mathbf f_u \mathbf s_u + n_u$, where $u=1,2,...U$ represents the $u$th user, The downlink SNR is defined as $P_t/\sigma_u^2$, where the $P_t$ is the transmission power and $\sigma_u^2$ is noise power spectrum.} 
A zero-forcing precoder is adopted and defined as $\tilde{\mathbf F} = \hat{\mathbf H}_b (\hat{\mathbf H}_b^H \hat{\mathbf H}_b)^{-1}$, where $\hat{\mathbf H}_b$ is the estimated uplink channels for $U$ users. The analysis assumes an ideal channel reciprocity and uses the estimated uplink channels as the downlink channels without considering a channel-reciprocity error.
For simplicity, the power allocation is not considered in simulating the sum rates; instead, each user's precoder uses matrix normalization to meet a total power constraint, i.e., $\mathbf f_u = \tilde{\mathbf f}_u/{\norm{\mathbf F}_F}$, where $ \tilde{\mathbf f}_{u}$ is the $u$th column of the precoder matrix $\tilde{\mathbf F}$. 
\bb{Performance comparison with the proposed methods is of two iterative algorithms including the OMP \cite{tropp2007signal} and AMP \cite{Donoho2009message} algorithms, and three deep-learning methods including a model-based network GM-LAMP \cite{Wei2021deep}, a model-free CNN-based network \cite{Ma2020data} and a sparse Bayesian learning unfolding network SBL-CE-Net \cite{gao2023deeplearning}. Note that the GM-LAMP has superior performance than LAMP \cite{Wei2021deep}, and LAMP has superior performance than LISTA \cite{Borgerding2017AMP}. The SBL-CE-Net method has demonstrated superior performance than a group of other methods including the MMV-LAMP \cite{ma2021modeldriven} and the SBL family \cite{Srivastava2019quasi, wang2017sparse}. Furthermore, results in \cite{ma2021modeldriven} and \cite{Srivastava2019quasi} have already demonstrated significant performance superiority than some other greedy algorithms such as SOMP, group-LASSO, and MMV-AMP \cite{gao2023deeplearning}.}

\subsection{ITRR and UTRR Performance Evaluations on Reconstructing Beamspace Channels}
\label{e_itrr_utrr_nmse}
In practice, beamspace channels are approximately sparse, containing most small-valued entries instead of zeros. It is of interest to examine the ITRR performance on reconstructing the testing dataset, which contains $2,000$ approximately beamspace channels, and compare with other widely-used iterative sparse reconstruction algorithms, including the OMP \cite{tropp2007signal} and AMP \cite{Donoho2009message} algorithms. 
Fig. 8 shows the results of channel estimation NMSE versus SNR.
The ITRR algorithm shows a better accuracy compared with the OMP algorithm over a wide SNR range, and has similar performance as the AMP algorithm.
Since the AMP and OMP algorithms have been successfully used in various compressive-sensing tasks, this result confirms the ITRR algorithm also has an competitive accuracy and can also be applied in similar applications. 
However, it is worth to mention the limitation of ITRR in the real-time applications that have low latency requirements.
First, the iterative mechanism requires sufficient computation time to achieve expected accuracy, but additional delay is undesirable.
\bb{Second, the top-K parameter as well as the regularization parameter $\rho$ in an ITRR algorithm needs to be manually evaluated and selected for better performance. Ideally, those parameters should be adaptively tuned according to the real channels to achieve best performance. Certain model-selection strategies such as cross validation can help to ensure overall performance for slowing changing environment. However, it is unrealistic to obtain the $K$ and $\rho$ that are ideal for all individual channel realizations.}

\begin{figure}[t]
\centering
\normalsize
\includegraphics[width=3.2in]{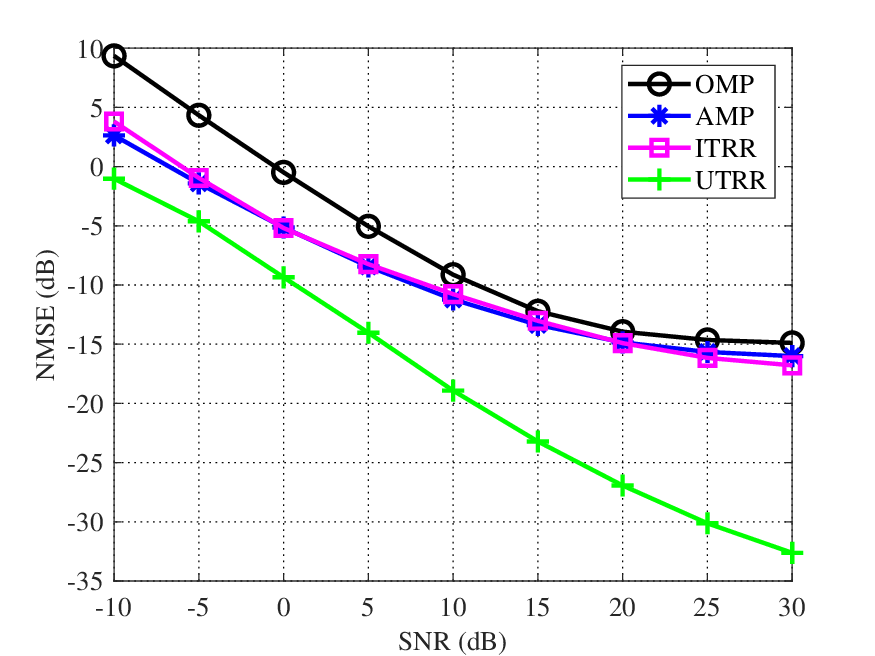}
\captionsetup{justification=centering}
\caption{Reconstruction NMSE of the iterative algorithms and the UTRR}
\label{nmse_compare_iter}
\end{figure}

The proposed deep-learning method can overcome the above limitations. The UTRR models used in our simulation have only $30$ unfolding layers, which is $20$ to $100$ times less than the number of ITRR iterations. Moreover, the trained models can be directly used to reconstruct a batch of channels simultaneously in parallel \bb{without tuning any parameters}. For example, is our simulation, we set the batch size as $128$, which means $64$ \footnote{The number is half of the batch size because we stack the real part and column part of a beamspace channels in column wise. Therefore, one beamspace channel is equivalent to two parallel real-valued vectors.} users' channels can be estimated simultaneously at one-time computation. 
It is of interest to evaluate the UTRR performance and compare with the ITRR algorithm over the same testing dataset.
Fig. \ref{nmse_compare_iter} includes the reconstruction NMSE by a trained UTRR model, where the top-$K$ parameter of the model is set to be $K=16$.
Compared with the iterative algorithms, the UTRR reduces the NMSE about $5$ dB at a low SNR, about $10$ dB at a moderate SNR, and about $15$ dB at a high SNR.

\subsection{UTRR Performance Evaluation on Reducing Computational Complexity}
\label{e_rcc}
Section \ref{sec_rcc} proposed the strategy of reducing computational complexity (RCC) for the UTRR model.
The computational complexity can be reduced by simply setting the parameter $K=0$ within the intermediate unfolding layers.
This section uses Table I to present the training time and reconstruction NMSE at $SNR=20$ dB for a UTRR model with the RCC configuration and compare with the results of the UTRR model without the RCC configuration.
In Table \ref{ccr_table}, the $\text{UTRR}_{K=16}$ without RCC is a UTRR model that contains $30$ layers of identical unfolding layers, where each layer has the top-$K$ parameter as $K=16$; the $\text{UTRR}_{K=16}$ with RCC is a UTRR model that sets the top-$K$ parameter $K=0$ for the $29$ layers of intermediate unfolding layers, while the top-$K$ parameter $K=16$ is for the last unfolding layer.
The UTRR model with a RCC and the UTRR model without a RCC have a similar NMSE but the former only costs about $50\%$ training time \footnote{The training was performed by a Nvidia RTX A6000 GPU.}.
Similar results have been observed in other SNR range and for other UTRR models having different the top-$K$ parameters. 
These results support our claim that the proposed RCC strategy does not degrade the model-prediction performance but can reduce the training time significantly.

\subsection{UTRR Performance Evaluation on Model-Averaging Ensemble}
\label{e_ensemble}
Section \ref{ensemble} proposed the model-averaging ensemble to improve the reconstruction accuracy. In a model-averaging-ensemble approach, \bb{several UTRR models having different top-$K$ parameters are trained individually, and then in the inference phase, channel estimations from different UTRR models are averaged as a final estimation.} 
Table \ref{ensamble_table_nmse} lists the NMSE results for different UTRR models having different top-$K$ parameters as well as the NMSE results for the final ensemble model UTRR-E at $SNR=10$ dB and $SNR=20$ dB. All the UTRR models with different top-$K$ parameters produced a similar NMSE, but by averaging the predictions from individual models, the UTRR-E produces a lower NMSE by around 4 dB.
Table \ref{ensamble_table_percent} measures the ratio of accurately reconstructed channels in the testing dataset, where an accurate reconstruction is counted when its normalized $\ell_2$ error is below a threshold, which is set as $Th=0.001$ for $SNR=10$ dB and $Th=0.0001$ for $SNR=20$ dB.
At both SNRs, Table \ref{ensamble_table_percent} shows the model-averaging ensemble improved the accurate reconstruction ratio by around $50\%$. 
The results of NMSE and the accurate reconstruction ratio from Table \ref{ensamble_table_nmse} and Table \ref{ensamble_table_percent} prove that the proposed model-averaging ensemble can improve the reconstruction accuracy significantly. Note that assembling multiple models that have the same the top-$K$ parameter would not produce such improvement.
In practical applications, since the computational complexity from the summation and average operations are negligible, the predictions from each trained model can be computed in parallel to avoid adding additional delays.

\begin{table}[!t]
\centering
\caption{Comparisons of the training time and reconstruction errors between a UTRR model with and without computational-complexity reduction (CCR). Here, the channel estimation SNR is $20$ dB, and the top-$K$ parameter of the UTRR model is $16$.}
\captionsetup{justification=raggedright}
\label{ccr_table}
\begin{tabular}{l l l}
\hline
\multicolumn{1}{l }{\textbf{Models}} & \begin{tabular}[c]{@{}l@{}} \textbf{NMSE} \end{tabular} & \begin{tabular}[c]{@{}l@{}}\textbf{Training Time} \end{tabular} \\ \hline
$\text{UTRR}_{K=16}$ w/o CCR  & $-26.78$ dB    & 548 min                                                              \\ 
$\text{UTRR}_{K=16}$ w/ CCR & $-26.93$ dB    & 255 min                                                              \\ \hline
\end{tabular}
\end{table}

\begin{table*}[!t]
\centering
\caption{Reconstruction NMSEs of different UTRR models and the ensembled model.}
\captionsetup{justification=raggedright}
\label{ensamble_table_nmse}
\begin{tabular}{l l l l l l}
\hline
 \multicolumn{1}{l }{\diagbox[width=10.5em]{\textbf{SNR}}{\textbf{NMSE(dB)}}{\textbf{Model}}}  & \begin{tabular}[c]{@{}l@{}} $\text{UTRR}_{K=0}$ \end{tabular} & \begin{tabular}[c]{@{}l@{}}$\text{UTRR}_{K=16}$\end{tabular} & \begin{tabular}[c]{@{}l@{}} $\text{UTRR}_{K=64}$\end{tabular} & \begin{tabular}[c]{@{}l@{}}$\text{UTRR}_{K=256}$\end{tabular} &  UTRR-E \\ \hline
$ M=128$, $SNR=10$ dB & $-18.92$  & $-18.92$   & $-19.00$   & $-18.77$   & $-22.06$   \\ 
$ M=128$, $SNR=20$ dB & $-26.97$  & $-26.93$   & $-26.73$   & $-26.94$   & $-30.18$   \\
$ M=64$, $SNR=20$ dB & $-21.89$  & $-21.22$   & $-21.72$   & $-21.45$   & $-24.51$  
 \\ \hline
\end{tabular}
\end{table*}

\begin{table*}[!t]
\centering
\caption{The accurate reconstruction ratios of different UTRR models and the ensembled model. Here, an accurate reconstruction is counted if its normalized $\ell_2$-error below the threshold $Th$.}
\captionsetup{justification=raggedright}
\label{ensamble_table_percent}
\begin{tabular}{l  l l l l l}
\hline
 \multicolumn{1}{l  }{\diagbox[width=15em]{\textbf{Threshold}}{\textbf{Ratio ($\%$)}}{\textbf{Model}}}  & \begin{tabular}[c]{@{}l@{}} $\text{UTRR}_{K=0}$ \end{tabular} & \begin{tabular}[c]{@{}l@{}}$\text{UTRR}_{K=16}$\end{tabular} & \begin{tabular}[c]{@{}l@{}} $\text{UTRR}_{K=64}$ \end{tabular} & \begin{tabular}[c]{@{}l@{}}$\text{UTRR}_{K=256}$\end{tabular} &  UTRR-E \\ \hline
$ M=128$, $Th=0.01$, $SNR=10$ dB & 39.1 &38.65 & 40.75 &34.6 & 85.15        \\ 
$ M=128$, $Th=0.001$, $SNR=20$ dB & 10.85     & 10.6    & 8.15    & 11.75   &  67.4 \\
$ M=64$, $Th=0.01$, $SNR=20$ dB & 85.8   & 81.4 & 85.2  & 83.1   & 95.6 
\\ \hline
\end{tabular}
\end{table*}

\subsection{Performance Comparison for Different Pilot Lengths}

\bb{The performance of UTRR models for pilot length $M=64$ are also added in  Table \ref{ensamble_table_nmse} and Table \ref{ensamble_table_percent} to show the pilot length influence.
Table \ref{ensamble_table_nmse} shows when using the pilot length of $M=64$, the NMSE performance degrades around 5 dB compared with the pilot length $M=128$.
Table \ref{ensamble_table_percent} shows the model-averaging ensemble improved the accurate reconstruction ratio by around $50\%$ when $M=128$. When $M=64$, the accurate reconstruction ratio improves by $10\%$--$15\%$ by model-averaging ensemble compared to the results from each individual model.
}

\begin{figure}[t]
\centering
\normalsize
\includegraphics[width=3.0in]{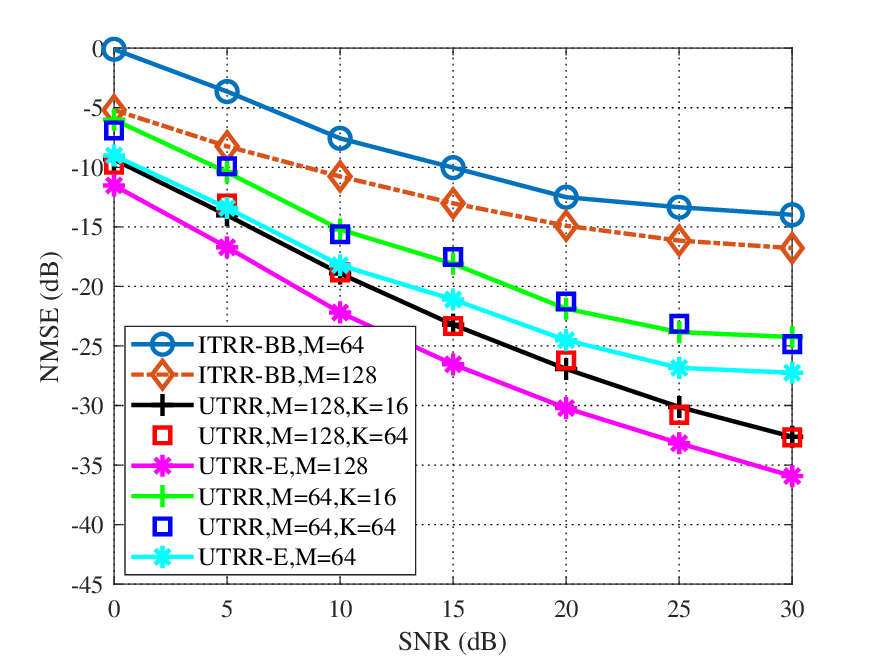}
\captionsetup{justification=centering}
\caption{Reconstruction NMSEs of deep-learning methods.}
\label{nmse_add_major_revision}
\end{figure}

\bb{Figure \ref{nmse_add_major_revision} compares the NMSE performance for the proposed approaches for the pilot lengths $M=128$ and $M=64$. Iterative aglorithms ITRR and ITRR-BB have similar accuracy performance, therefore, we only demonstrate ITRR-BB accuracy in this figure. It shows that more pilots lead to higher accuracy for all approaches. Compared with ITRR-BB, the unfolded networks UTRR and UTRR-E using the same pilot length achieved higher accuracy.
For UTRR and UTRR-E, more pilots lead to more noticeable NMSE reduction at high SNR range between $20$--$30$ dB.}

\subsection{Comparisons of UTRR Performance With Other Deep-Learning Methods}
\label{e_utrr_compare}

\bb{Next, we compare the UTRR-E performance with the GM-LAMP model \cite{Wei2021deep}, the CNN-based model \cite{Ma2020data}, and the SBL-CE-Net \cite{gao2023deeplearning}.  
Fig. 9 plots the channel estimation NMSE over the SNR in the range of $[-10, 25]$ dB. 
Compared with the GM-LAMP and the CNN-based model, the proposed UTRR-E achieved lower NMSEs for the entire SNR range. 
The CNN-based model has a slightly lower NMSE than the GM-LAMP for the low SNR range of $[-10, 5]$ dB, but at a high $SNR \in[10, 25]$ dB, the NMSE of the CNN-based model stuck at around $-15$ dB while other methods start showing noticeable advantages. For example, at the high $SNR = 25$ dB, the SBL-CE-Net has the NMSE around $-25$ dB, the GM-LAMP achieves the NMSE around $-26$ dB, and the proposed UTRR-E achieves the NMSE below $-33$ dB.}

As a widely used metric to evaluate the channel estimation accuracy, the NMSE itself is insufficient to measure the impact of imperfect channel estimation on the communication system.
Because the estimated channels from two different channel estimation approaches can have the same NMSE but have significantly distinct channel realizations. Different channel realizations having different values in each antenna entry can lead to different beamforming directions and thus cause a significantly different impact on the achievable rate of the massive MIMO system.

\begin{figure}[t]
\centering
\normalsize
\includegraphics[width=3.0in]{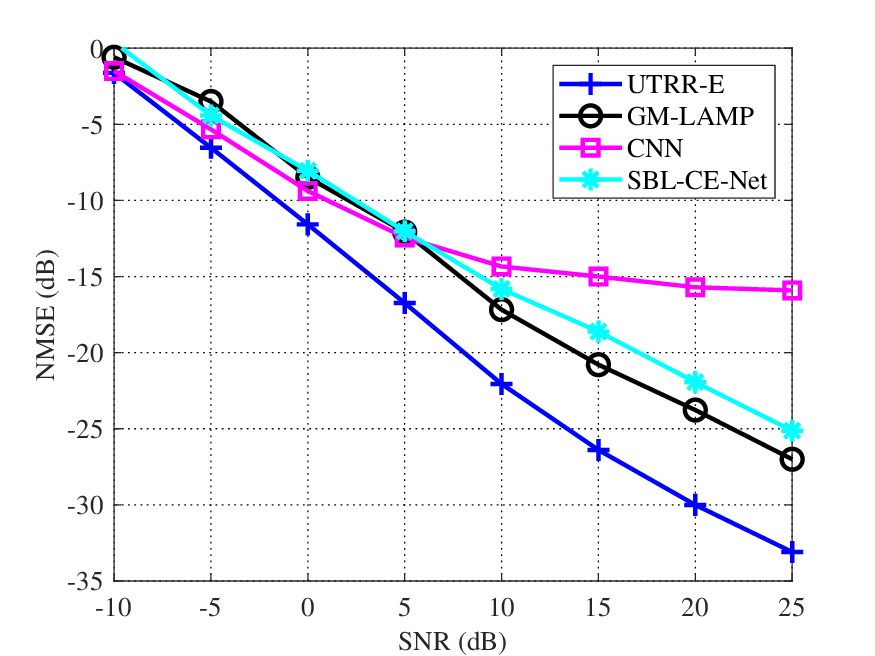}
\captionsetup{justification=centering}
\caption{Reconstruction NMSEs of deep-learning methods.}
\end{figure}

Figure 10 plots the sum rate of using the estimated beamspace channels by the proposed UTRR-E and also plots the sum rate of using the Gaussian noise-contaminated CSI. The x-axis is the channel estimation NMSE to ensure the sum rates are computed at the same NMSE for two types of imperfect CSI. 
Large gaps exist between the sum rates of using the UTRR-E estimated CSI and using the Gaussian noise contaminated CSI. 
This result affirms that different channel estimation schemes can have the same NMSE but have a significantly different impact on the sum rate. Therefore, the sum rate is a necessary metric to access the direct impact of channel estimation schemes on a communication system.

\begin{figure}[t]
\centering
\normalsize
\includegraphics[width=3.2in]{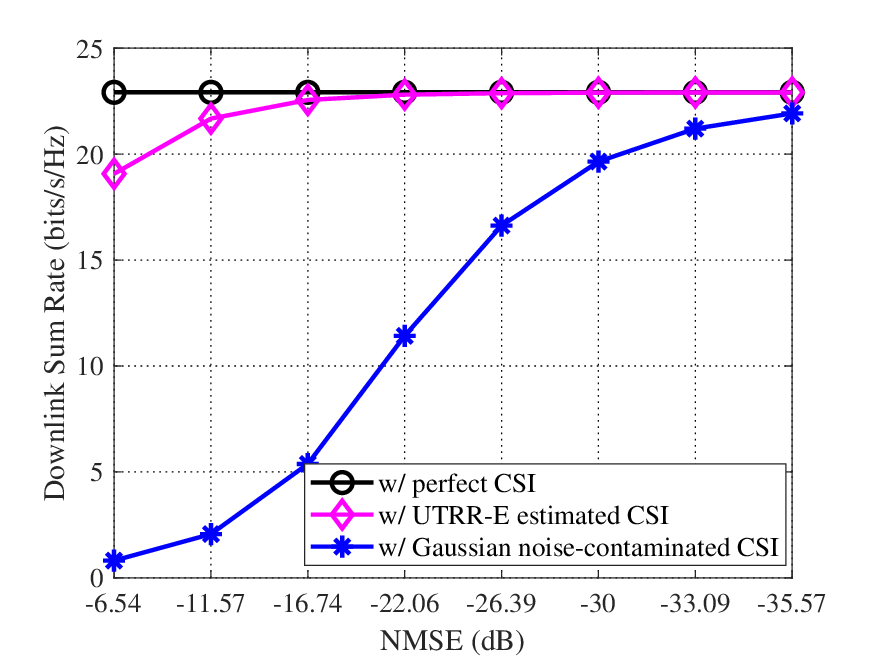}
\captionsetup{justification=centering}
\caption{Downlink sum rate at $SNR_{dl} = 15$ dB versus different channel estimation NMSE}
\end{figure}

\begin{figure}[t]
\centering
\normalsize
\includegraphics[width=3.2in]{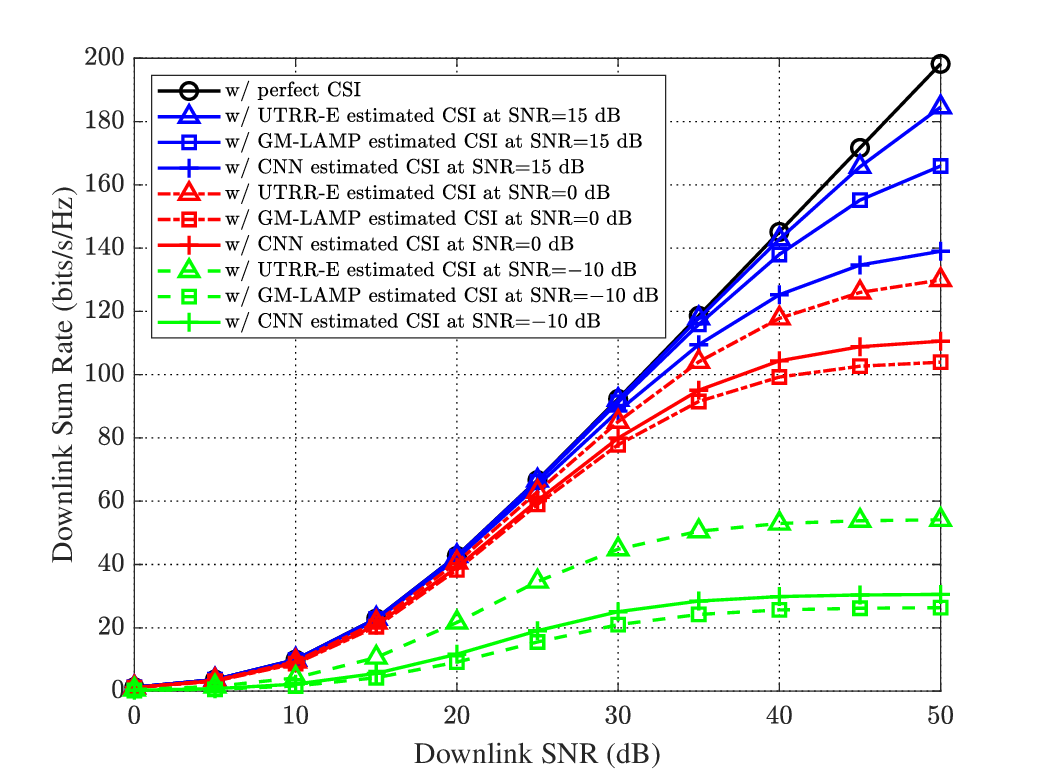}
\captionsetup{justification=centering}
\caption{Downlink Sum rate versus downlink SNR when uplink channels are estimated at $SNR = -10, 0, 15 $ dB}
\end{figure}

\begin{figure}[t]
\centering
\normalsize
\includegraphics[width=3.2in]{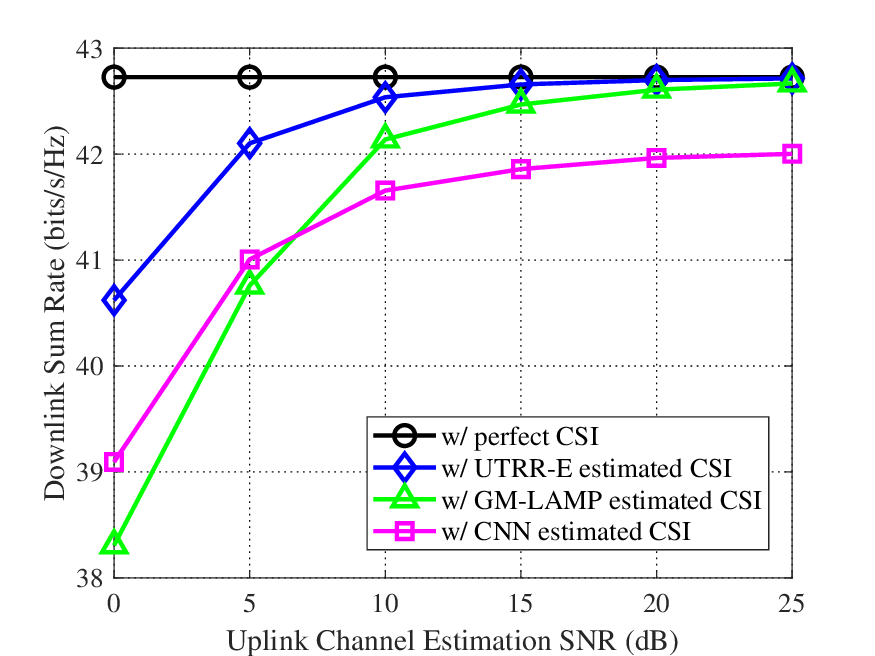}
\caption{Downlink sum rate at $SNR_{dl} = 20$ dB versus uplink channel estimation SNR}
\end{figure}

Figure 11 compares the downlink sum rates versus the transmit SNR for different deep-learning channel-estimation methods, where the ideal channel reciprocity is assumed and the uplink channels were estimated at $SNR = 15, 0, -10$ dB.
When the uplink channels are estimated at a high SNR, the channel estimation errors are small enough. Thus, the downlink sum rates are dominated by the downlink SNR. Therefore, all three deep-learning methods show little difference in the sum rates. 
For example, for the uplink channels are estimated when the channel estimation SNR is $15$ dB, the average channel estimation errors are $-26.39$ dB, $-20.79$ dB, and $-15$ dB for the UTRR-E, the GM-LAMP and the CNN-based model, respectively; their sum rates are $143$, $138$, and $125$ bits/s/Hz for the downlink $SNR=40$ dB, while the sum rate with perfect CSI is $145$ bits/s/Hz that only has little difference with the sum rates using imperfect CSI. 
The uplink channel-estimation errors increase when the uplink SNR decreases, and thus the downlink sum rates degrade more obviously because of the imperfect CSI. 
As shown in the sum-rate curves when the channel estimation $SNR= -10$ dB, larger gaps exist between the sum rates using imperfect CSI and the sum rate using perfect CSI. 
For all the cases of different channel estimation SNRs, the proposed UTRR-E model achieved higher sum rate over the entire downlink SNR range. 
Fig. 12 shows the impact of the channel estimation SNR on the downlink sum rate, where the downlink transmit SNR is $SNR_{dl} = 20$ dB. 
When the channel estimation SNR increases, the sum rates of all three deep-learning methods increase and gradually approach their maximum sum rates. 
\bb{The optimal sum rate with perfect CSI is 43 bits/s/Hz, which is drawn as a straight line for comparison. 
The UTRR-E achieves a 41 bits/s/Hz sum rate, which is $95 \%$ of the optimal sum rate at $SNR = 0$ dB; at $SNR = 10$ dB, the UTRR-E achieves the optimal sum rate.}
For the entire uplink channel estimation SNR range, the UTRR-E shows higher sum rates than the other two deep-learning methods.
The sum rate results of Fig. 11 and Fig. 12 conclude that the proposed UTRR models with model-averaging ensemble can estimate the channels more accurately and provide higher sum rate than the state-of-the-art other deep-learning methods.

\subsection{Computational Complexity}
\bb{The computational complexity of the ITRR algorithm is $O(TMN^2)$, where $T$ represents the number of iterations before convergence. The forward-pass computational complexity of a  UTRR network is $O(LMN^2)$, where $L$ represents the number of layers of the decoder. Although the computational complexities of the ITRR and the UTRR decoder appear similar, the number of layers $L$ is usually much smaller than the number of iterations $I$. Additionally, the sorting operation of finding largest $K$ elements has computational complexity $O(N log K)$. ITRR algorithm requires to perform sorting operation in every iteration while the UTRR with CRR only requires once in the last layer of the network.
By contrast, the GM-LAMP network are constructed on the AMP algorithm, the computational complexity of the AMP algorithm and the GM-LAMP network is the same, i.e., $O(TMN)$ \cite{Wei2021deep}.
The computational complexity of the CNN-based model is $O(NM)$ for each fully connected layer and $O(N k^2 \Sigma_{l=1}^{N_{re}}n_{l-1}n_l)$ for each CNN layer, where k is the side length of the convolutional filters, $n_{l-1}$ and $n_l$ denote the numbers of input and output feature maps of the $l$the convolutional layer, $1 \le l \le N_{re}$ \cite{Ma2020data}; seven CNN layers were used with the kernel size $ 3 \times 3$ \cite{Ma2020data}.  
The computational complexity of the OMP algorithm can be represented by $O(SMN) + O(S^3M)$, where $S$ is the sparsity level of the beamspace channel vector \cite{Venugopal2017channel}.
The computational complexity for SBL family is $O(TMN^2)$. Note that the computational complexity is quite high because calculating matrix inverse is required for two times per iteration, and the computational complexity for matrix inverse is $O(M^3)$. 
The proposed SBL-CE-Net introduces extra computational complexity by the 3D convolutional layers is $O(N N_F F_S^3)$, where $N_F$ is the number of filters NF filters and the filter size is $F_S$ \cite{gao2023deeplearning}.}

\section{Conclusion}

This paper presented an iterative algorithm, i.e., the ITRR, and then unfolded it into a deep-learning model, i.e., the UTRR, for the beamspace channel estimation in mmWave massive MIMO systems. 
The ITRR algorithm has interpretable structure  and shows competitive performance, while the practical implemetation is challenging due to high computational complexity and suboptimal parameter tunning.
By unfolding the ITRR into a learnable algorithm which can be trained offline, the UTRR addressed the practical implementation challenges with faster, more robust and accurate sparse reconstructions. Furthermore, this paper proposed the model-averaging ensemble to improve the model prediction accuracy. 
Numerical results conclude that the proposed learning scheme can reconstruct beamspace channels with a higher accuracy and thus prompts the mmWave massive MIMO system to achieve larger sum rates. 
This work demonstrated a complete design of leveraging traditional model-driven optimizations to develop a deep-learning method for an applied problem. 
The proposed UTRR models can be applied to other sparse reconstruction applications.

\bibliographystyle{IEEEtran}

\bibliography{IEEEabrv,Nov2022TCOM2.bib}

\begin{thebibliography}{10}
\providecommand{\url}[1]{#1}
\csname url@samestyle\endcsname
\providecommand{\newblock}{\relax}
\providecommand{\bibinfo}[2]{#2}
\providecommand{\BIBentrySTDinterwordspacing}{\spaceskip=0pt\relax}
\providecommand{\BIBentryALTinterwordstretchfactor}{4}
\providecommand{\BIBentryALTinterwordspacing}{\spaceskip=\fontdimen2\font plus
\BIBentryALTinterwordstretchfactor\fontdimen3\font minus
  \fontdimen4\font\relax}
\providecommand{\BIBforeignlanguage}[2]{{%
\expandafter\ifx\csname l@#1\endcsname\relax
\typeout{** WARNING: IEEEtran.bst: No hyphenation pattern has been}%
\typeout{** loaded for the language `#1'. Using the pattern for}%
\typeout{** the default language instead.}%
\else
\language=\csname l@#1\endcsname
\fi
#2}}
\providecommand{\BIBdecl}{\relax}
\BIBdecl

\bibitem{heath2016an}
R.~W. {Heath Jr.}, N.~González-Prelcic, S.~Rangan, W.~Roh, and A.~M. Sayeed,
  ``An overview of signal processing techniques for millimeter wave {MIMO}
  systems,'' \emph{IEEE J. Sel. Topics Signal Process.}, vol.~10, no.~3, pp.
  436--453, Apr. 2016.

\bibitem{Busari2018millimeter}
S.~A. Busari, K.~M.~S. Huq, S.~Mumtaz, L.~Dai, and J.~Rodriguez,
  ``Millimeter-wave massive {MIMO} communication for future wireless systems: A
  survey,'' \emph{IEEE Commun. Surveys Tut.}, vol.~20, no.~2, pp. 836--869,
  Second Quart. 2018.

\bibitem{Berger2010application}
C.~R. {Berger}, Z.~{Wang}, J.~{Huang}, and S.~{Zhou}, ``Application of
  compressive sensing to sparse channel estimation,'' \emph{IEEE Commun. Mag.},
  vol.~48, no.~11, pp. 164--174, Nov. 2010.

\bibitem{yonina2012CS}
Y.~C. Eldar and G.~Kutyniok, \emph{Compressed Sensing: Theory and
  Applications}.\hskip 1em plus 0.5em minus 0.4em\relax New York, NY, USA:
  Cambridge University Press, 2012.

\bibitem{yonina2015sampling}
Y.~C. Eldar, \emph{Sampling Theory: Beyond Bandlimited Systems}.\hskip 1em plus
  0.5em minus 0.4em\relax New York, NY, USA: Cambridge University Press, 2015.

\bibitem{gao2015spatially}
Z.~Gao, L.~Dai, Z.~Wang, and S.~Chen, ``Spatially common sparsity based
  adaptive channel estimation and feedback for {FDD} massive {MIMO},''
  \emph{IEEE Trans. Signal Process.}, vol.~63, no.~23, pp. 6169--6183, Dec.
  2015.

\bibitem{gao2016structured}
Z.~Gao, L.~Dai, W.~Dai, B.~Shim, and Z.~Wang, ``Structured compressive
  sensing-based spatio-temporal joint channel estimation for {FDD} massive
  {MIMO},'' \emph{IEEE Trans. Commun.}, vol.~64, no.~2, pp. 601--617, Feb.
  2016.

\bibitem{eltayeb2014compressive}
M.~E. Eltayeb, T.~Y. Al-Naffouri, and H.~R. Bahrami, ``Compressive sensing for
  feedback reduction in {MIMO} broadcast channels,'' \emph{IEEE Trans.
  Commun.}, vol.~62, no.~9, pp. 3209--3222, Sept. 2014.

\bibitem{choi2015downlink}
J.~W. Choi, B.~Shim, and S.~Chang, ``Downlink pilot reduction for massive
  {MIMO} systems via compressed sensing,'' \emph{IEEE Commun. Lett.}, vol.~19,
  no.~11, pp. 1889--1892, Nov. 2015.

\bibitem{tseng2016enhanced}
C.-C. Tseng, J.-Y. Wu, and T.-S. Lee, ``Enhanced compressive downlink {CSI}
  recovery for {FDD} massive {MIMO} systems using weighted block
  $l_1$-minimization,'' \emph{IEEE Trans. Commun.}, vol.~64, no.~3, pp.
  1055--1067, Mar. 2016.

\bibitem{shen2016joint}
W.~Shen, L.~Dai, Y.~Shi, B.~Shim, and Z.~Wang, ``Joint channel training and
  feedback for {FDD} massive {MIMO} systems,'' \emph{IEEE Trans. Veh.
  Technol.}, vol.~65, no.~10, pp. 8762--8767, Oct. 2016.

\bibitem{rao2014distributed}
X.~Rao and V.~K.~N. Lau, ``Distributed compressive {CSIT} estimation and
  feedback for {FDD} multi-user massive {MIMO} systems,'' \emph{IEEE Trans.
  Signal Process.}, vol.~62, no.~12, pp. 3261--3271, June 2014.

\bibitem{liu2017closed}
A.~Liu, F.~Zhu, and V.~K.~N. Lau, ``Closed-loop autonomous pilot and
  compressive {CSIT} feedback resource adaptation in multi-user {FDD} massive
  {MIMO} systems,'' \emph{IEEE Trans. Signal Process.}, vol.~65, no.~1, pp.
  173--183, Jan. 2017.

\bibitem{wu2021nonconvex}
P.~Wu and J.~Cheng, ``Nonconvex regularized gradient projection sparse
  reconstruction for massive {MIMO} channel estimation,'' \emph{IEEE Trans.
  Commun}, vol.~69, no.~11, pp. 7722--7735, Nov. 2021.

\bibitem{gao2017reliable}
X.~Gao, L.~Dai, S.~Han, C.~I, and X.~Wang, ``Reliable beamspace channel
  estimation for millimeter-wave massive {MIMO} systems with lens antenna
  array,'' \emph{IEEE Trans. Wireless Commun.}, vol.~16, no.~9, pp. 6010--6021,
  Sept. 2017.

\bibitem{gao2016channel_letter}
Z.~Gao, C.~Hu, L.~Dai, and Z.~Wang, ``Channel estimation for millimeter-wave
  massive {MIMO} with hybrid precoding over frequency-selective fading
  channels,'' \emph{IEEE Commun. Lett.}, vol.~20, no.~6, pp. 1259--1262, June
  2016.

\bibitem{Mo2018}
P.~S. Jianhua~Mo and R.~W. {Heath Jr.}, ``Channel estimation in broadband
  millimeter wave {MIMO} systems with few-bit {ADCs},'' \emph{IEEE Trans.
  Signal Process.}, vol.~66, no.~5, pp. 1141--1154, Mar. 2018.

\bibitem{tropp2007signal}
J.~A. Tropp and A.~C. Gilbert, ``Signal recovery from random measurements via
  orthogonal matching pursuit,'' \emph{IEEE Trans. Inform. Theory}, vol.~53,
  no.~12, pp. 4655--4666, Dec. 2007.

\bibitem{Venugopal2017}
K.~Venugopal, A.~Alkhateeb, N.~G. Prelcic, and R.~W. {Heath Jr.}, ``Channel
  estimation for hybrid architecture-based wideband millimeter wave systems,''
  \emph{IEEE J. Sel. Areas Commun.}, vol.~35, no.~9, pp. 1996--2009, Sept.
  2017.

\bibitem{coma2018channel}
J.~P. Gonz\'alez-Coma, J.~Rodr\'iguez-Fern\'andez, N.~Gonz\'alez-Prelcic,
  L.~Castedo, and R.~W. {Heath Jr.}, ``Channel estimation and hybrid precoding
  for frequency selective multiuser {mmWave} {MIMO} systems,'' \emph{IEEE J.
  Sel. Topics Signal Process.}, vol.~12, no.~2, pp. 353--367, May 2018.

\bibitem{rod2018frequency}
J.~{Rodr\'iguez-Fern\'andez}, N.~{Gonz\'alez-Prelcic}, K.~{Venugopal}, and
  R.~W. {Heath Jr.}, ``Frequency-domain compressive channel estimation for
  frequency-selective hybrid millimeter wave {MIMO} systems,'' \emph{IEEE
  Trans. Wireless Commun.}, vol.~17, no.~5, pp. 2946--2960, May 2018.

\bibitem{Srivastava2019quasi}
S.~Srivastava, A.~Mishra, A.~Rajoriya, A.~K. Jagannatham, and G.~Ascheid,
  ``Quasi-static and time-selective channel estimation for block-sparse
  millimeter wave hybrid {MIMO} systems: Sparse bayesian learning ({SBL}) based
  approaches,'' \emph{IEEE Trans. Signal Process.}, vol.~67, no.~5, pp.
  1251--1266, Mar. 2019.

\bibitem{wang2017sparse}
P.~Wang, M.~Pajovic, P.~V. Orlik, T.~Koike-Akino, K.~J. Kim, and J.~Fang,
  ``Sparse channel estimation in millimeter wave communications: Exploiting
  joint {AoD-AoA} angular spread,'' in \emph{Proc. IEEE Int. Conf. Commun.
  (ICC)}, Taipei, Taiwan, May 2017, pp. 1--6.

\bibitem{joseph2017noniterative}
G.~Joseph and C.~R. Murthy, ``Channel estimation for hybrid architecture based
  wideband millimeter wave systems,'' \emph{IEEE Trans. Signal Process.},
  vol.~65, no.~20, pp. 5510--5525, Otc. 2017.

\bibitem{Donoho2009message}
D.~L. Donoho, A.~Maleki, and A.~Montanari, ``Message passing algorithms for
  compressed sensing,'' \emph{Natl. Acad. Sci.}, vol. 106, no.~45, pp.
  18\,914--18\,919, 2009.

\bibitem{gao2023deeplearning}
J.~Gao, C.~Zhong, G.~Y. Li, and A.~B. J.~B.~Soriaga, ``Deep learning-based
  channel estimation for wideband hybrid mmwave massive {MIMO},'' \emph{IEEE
  Trans. Commun.}, vol.~71, no. 106, pp. 3679--3693, June 2023.

\bibitem{chen2021hybrid}
Y.~Chen, Y.~Xiong, D.~Chen, T.~Jiang, S.~X. Ng, and L.~Hanzo, ``Hybrid
  precoding for wideband millimeter wave {MIMO} systems in the face of beam
  squint,'' \emph{IEEE Trans. Wireless Commun.}, vol.~20, no.~1, pp.
  1847--1859, Mar. 2021.

\bibitem{he2018deep}
H.~He, C.-K. Wen, S.~Jin, and G.~Y. Li, ``Deep learning-based channel
  estimation for beamspace {mmWave} massive {MIMO} systems,'' \emph{IEEE
  Wireless Commun. Lett.}, vol.~7, no.~5, pp. 852--855, Oct. 2018.

\bibitem{chun2019deep}
C.-J. Chun, J.-M. Kang, and I.-M. Kim, ``Deep learning-based channel estimation
  for massive {MIMO} systems,'' \emph{IEEE Commun. Lett.}, vol.~8, no.~4, pp.
  1228--1231, Aug. 2019.

\bibitem{wen2018deep}
C.~K. Wen, W.~T. Shih, and S.~Jin, ``Deep learning for massive {MIMO} {CSI}
  feedback,'' \emph{IEEE Wireless Commun. Lett.}, vol.~7, no.~5, pp. 748--751,
  Oct. 2018.

\bibitem{wang2018deep}
T.~Wang, C.~Wen, S.~Jin, and G.~Y. Li, ``Deep learning-based {CSI} feedback
  approach for time-varying massive {MIMO} channels,'' \emph{IEEE Wireless
  Commun. Lett.}, vol.~8, no.~2, pp. 416--419, Apr. 2019.

\bibitem{yang2019deep}
Q.~{Yang}, M.~B. {Mashhadi}, and D.~{Gündüz}, ``Deep convolutional
  compression for massive {MIMO} {CSI} feedback,'' in \emph{2019 IEEE 29th
  International Workshop on Machine Learning for Signal Processing (MLSP)},
  Oct. 2019, pp. 1--6.

\bibitem{Ma2020data}
X.~Ma and Z.~Gao, ``Data-driven deep learning to design pilot and channel
  estimator for massive {MIMO},'' \emph{IEEE Trans. Veh. Technol.}, vol.~69,
  no.~5, pp. 5677--5682, May 2020.

\bibitem{Mashhadi2021pruning}
M.~B. Mashhadi and D.~G\"{u}nd\"{u}z, ``Pruning the pilots: Deep learning-based
  pilot design and channel estimation for {MIMO-OFDM} systems,'' \emph{IEEE
  Trans. Wireless Commun}, vol.~20, no.~10, pp. 6315--6328, Oct. 2021.

\bibitem{Shlezinger2022model}
N.~Shlezinger, Y.~C. Eldar, and S.~P. Boyd, ``Model-based deep learning: On the
  intersection of deep learning and optimization,'' \emph{IEEE Access},
  vol.~10, pp. 115\,384--115\,398, Nov. 2022.

\bibitem{Shlezinger2023model}
N.~Shlezinger, J.~Whang, Y.~C. Eldar, and A.~G. Dimakis, ``Model-based deep
  learning,'' \emph{Proceedings of the IEEE}, vol. 111, no.~5, pp. 465--499,
  May 2023.

\bibitem{he2019model}
H.~{He}, S.~{Jin}, C.~{Wen}, F.~{Gao}, G.~Y. {Li}, and Z.~{Xu}, ``Model-driven
  deep learning for physical layer communications,'' \emph{IEEE Wireless
  Commun.}, vol.~26, no.~5, pp. 77--83, Oct. 2019.

\bibitem{hershey2014deep}
J.~Hershey, J.~L. Roux, and F.~Weninger, ``Deep unfolding: {Model-based}
  inspiration of novel deep architectures,'' \emph{Mitsubishi Electric Research
  Laboratories, Tech. Rep. TR2014-117}, Aug. 2014.

\bibitem{Monga2021algorithm}
V.~Monga, Y.~Li, and Y.~C. Eldar, ``Algorithm unrolling: {Interpretable},
  efficient deep learning for signal and image processing,'' \emph{IEEE Signal
  Process. Mag.}, vol.~38, no.~2, pp. 18--44, Mar. 2021.

\bibitem{Gregor2010Learning}
K.~Gregor and Y.~LeCun, ``Learning fast approximations of sparse coding,'' in
  \emph{Proceedings of the 27th International Conference on Machine Learning
  (ICML)}, 2010, pp. 399--406.

\bibitem{metzler2017learned}
C.~Metzler, A.~Mousavi, and R.~Baraniuk, ``Learned {D-AMP}: Principled neural
  network based compressive image recovery,'' in \emph{Advances in Neural
  Information Processing Systems (NeurIPS)}, Long Beach, CA, USA, 2017, pp.
  1772--1783.

\bibitem{Ma2021model}
X.~Ma, Z.~Gao, F.~Gao, and M.~D. Renzo, ``Model-driven deep learning based
  channel estimation and feedback for millimeter-wave massive hybrid {MIMO}
  systems,'' \emph{IEEE J. Sel. Areas Commun.}, vol.~39, no.~8, pp. 2388--2406,
  2021.

\bibitem{wu2022deep}
P.~Wu and J.~Cheng, ``Deep unfolding basis pursuit: Improving sparse channel
  reconstruction via data-driven measurement matrices,'' \emph{IEEE Trans.
  Wireless Commun.}, vol.~21, no.~10, pp. 8090--8150, Oct. 2022.

\bibitem{Tono2017efficient}
K.~Tono, A.~Takeda, and J.~Gotoh., ``Efficient {DC} algorithm for constrained
  sparse optimization,'' 2017. [Online]. Available: arXiv:1701.08498.

\bibitem{Wei2021deep}
X.~Wei, C.~Hu, and L.~Dai, ``Deep learning for beamspace channel estimation in
  millimeter-wave massive {MIMO} systems,'' \emph{IEEE Trans. Commun.},
  vol.~69, no.~1, pp. 182--193, Jan. 2021.

\bibitem{brady2013beamspace}
J.~Brady, N.~Behdad, and A.~M. Sayeed, ``Beamspace {MIMO} for millimeter-wave
  communications: {System} architecture, modeling, analysis, and
  measurements,'' \emph{IEEE Trans. Antennas Propag.}, vol.~61, no.~7, pp.
  3814--3827, July 2013.

\bibitem{chen2019channel}
Y.~Chen, D.~Chen, and L.~Hanzo, ``Channel-covariance and angle-of-departure
  aided hybrid precoding for wideband multiuser millimeter wave {MIMO}
  systems,'' \emph{IEEE Trans. Commun.}, vol.~67, no.~12, pp. 8315--8327, Dec.
  2019.

\bibitem{Peng2021thesis}
P.~Wu, ``Sparse channel estimation for massive {MIMO} via deep compressive
  sensing,'' Ph.D. dissertation, University of British Columbia, 2021.

\bibitem{Thi2018DC}
H.~A.~L. Thi and T.~P. Dinh, ``{DC} programming and {DCA}: Thirty years of
  developments,'' \emph{Math. Program.}, vol. 169, pp. 5--68, 2018.

\bibitem{Parikh2014proximal}
N.~Parikh and S.~P. Boyd, ``Proximal algorithms,'' \emph{Found. Trends Optim.},
  vol.~1, no.~3, pp. 123--231, 2014.

\bibitem{he2016deep}
K.~He, X.~Zhang, S.~Ren, and J.~Sun, ``Deep residual learning for image
  recognition,'' in \emph{Proceedings of the IEEE conference on computer vision
  and pattern recognition (CVPR)}, Las Vegas, NV, USA, June 2016, pp. 770--778.

\bibitem{Sollich1995learning}
P.~Sollich and A.~Krogh, ``Learning with ensembles: How over-fitting can be
  useful,'' in \emph{Advances in Neural Information Processing Systems (NIPS)},
  1995, pp. 190--196.

\bibitem{ensemble1999liu}
Y.~Liu and X.~Yao, ``Ensemble learning via negative correlation,'' \emph{Neural
  Networks}, vol.~12, no.~10, pp. 1399--1404, Dec. 1999.

\bibitem{improving2022jose}
D.~M. Jose, A.~M. Vincent, and G.~S. Dwarakish, ``Improving multiple model
  ensemble predictions of daily precipitation and temperature through machine
  learning techniques,'' \emph{Scientific Reports}, vol.~12, no. 4678, Mar.
  2022.

\bibitem{Borgerding2017AMP}
M.~Borgerding, P.~Schniter, and S.~Rangan, ``{AMP}-inspired deep net-works for
  sparse linear inverse problems,'' \emph{IEEE Trans. Signal Process.},
  vol.~65, no.~16, pp. 4293--4308, Aug. 2017.

\bibitem{ma2021modeldriven}
X.~Ma, Z.~Gao, F.~Gao, and M.~D. Renzo, ``Model-driven deep learning based
  channel estimation and feedback for millimeter-wave massive hybrid {MIMO}
  systems,'' \emph{IEEE J. Sel. Areas Commun.}, vol.~39, no.~8, pp. 2388--2406,
  Aug. 2021.

\bibitem{Venugopal2017channel}
K.~Venugopal, A.~Alkhateeb, N.~G. Prelcic, and R.~W. Heath, ``Channel
  estimation for hybrid architecture based wideband millimeter wave systems,''
  \emph{IEEE J. Sel. Areas Commun.}, vol.~35, no.~9, pp. 1996--2009, Sept.
  2017.

\end{thebibliography}


\begin{IEEEbiography}[{\includegraphics[width=1.05in,height=1.25in,clip,keepaspectratio]{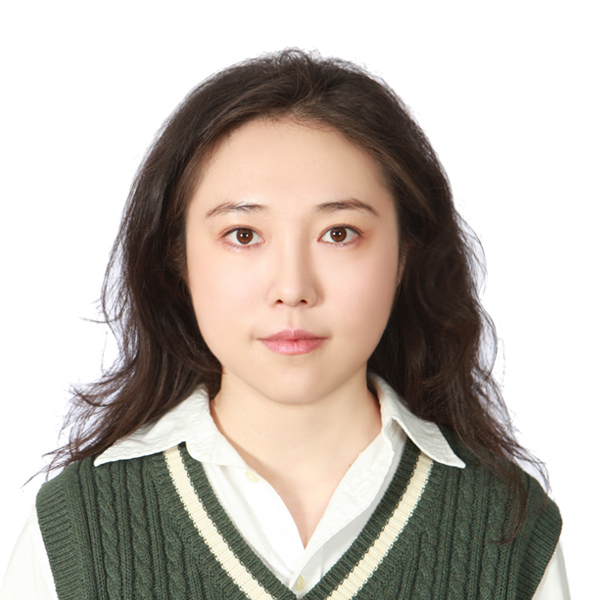}}]{Pengxia Wu}
received the B.S. degree in communication engineering in 2013 and the M.S degree in communication and information system in 2016 from Jilin University, Changchun, China. She received the Ph.D. degree in electrical engineering from the University of British Columbia, School of Engineering, Kelowna, BC, Canada, in 2022. She is currently a research scientist at Rockwell Automation, Milwaukee, WI, USA. Her research interests include wireless communication, robotics, and artificial intelligence in industrial automation.
\end{IEEEbiography}
\vskip -0.8\baselineskip plus -1fil 

\begin{IEEEbiography}[{\includegraphics[width=1in,height=1.25in,clip,keepaspectratio]{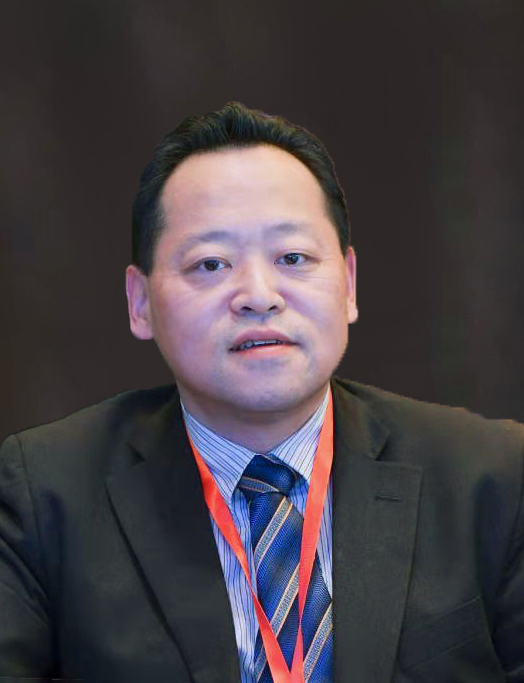}}]{Julian Cheng}
(S’96–M’04–SM’13–F’23) received a B.Eng. degree (Hons.) in electrical engineering from the University of Victoria, Victoria, BC, Canada, in 1995, the M.Sc.(Eng.) degree in mathematics and engineering from Queen’s University, Kingston, ON, Canada, in 1997, and a Ph.D. in electrical engineering from the University of Alberta, Edmonton, AB, Canada, in 2003. He is a Full Professor in the School of Engineering, Faculty of Applied Science, The University of British Columbia, Kelowna, BC, Canada. He was with Bell Northern Research and NORTEL Networks. His research interests include machine learning and deep learning for wireless communications, wireless optical technology, and quantum communications. He is a Registered Professional Engineer in British Columbia, Canada. He served as the President of the Canadian Society of Information Theory (2017-2021). He was the Co-Chair of the 12th Canadian Workshop on Information Theory in 2011, the 28th Biennial Symposium on Communications in 2016, and the General Co-Chair of the 2021 and 2024 IEEE Communication Theory Workshop. He is the Chair of the Radio Communications Technical Committee of the IEEE Communications Society. He was a past Associate Editor of the IEEE TRANSACTIONS ON COMMUNICATIONS, the IEEE TRANSACTIONS ON WIRELESS COMMUNICATIONS, the IEEE COMMUNICATIONS LETTERS, and the IEEE ACCESS, as well as an Area Editor for the IEEE TRANSACTIONS ON COMMUNICATIONS. Dr. Cheng served as a Guest Editor for Special Issues of the IEEE JOURNAL ON SELECTED AREAS IN COMMUNICATIONS on Optical Wireless Communications and Positioning and Sensing Over Wireless Networks.
\end{IEEEbiography}
\vskip -1\baselineskip plus -1fil 

\begin{IEEEbiography}[{\includegraphics[width=1.1in,height=1.25in,clip,keepaspectratio]{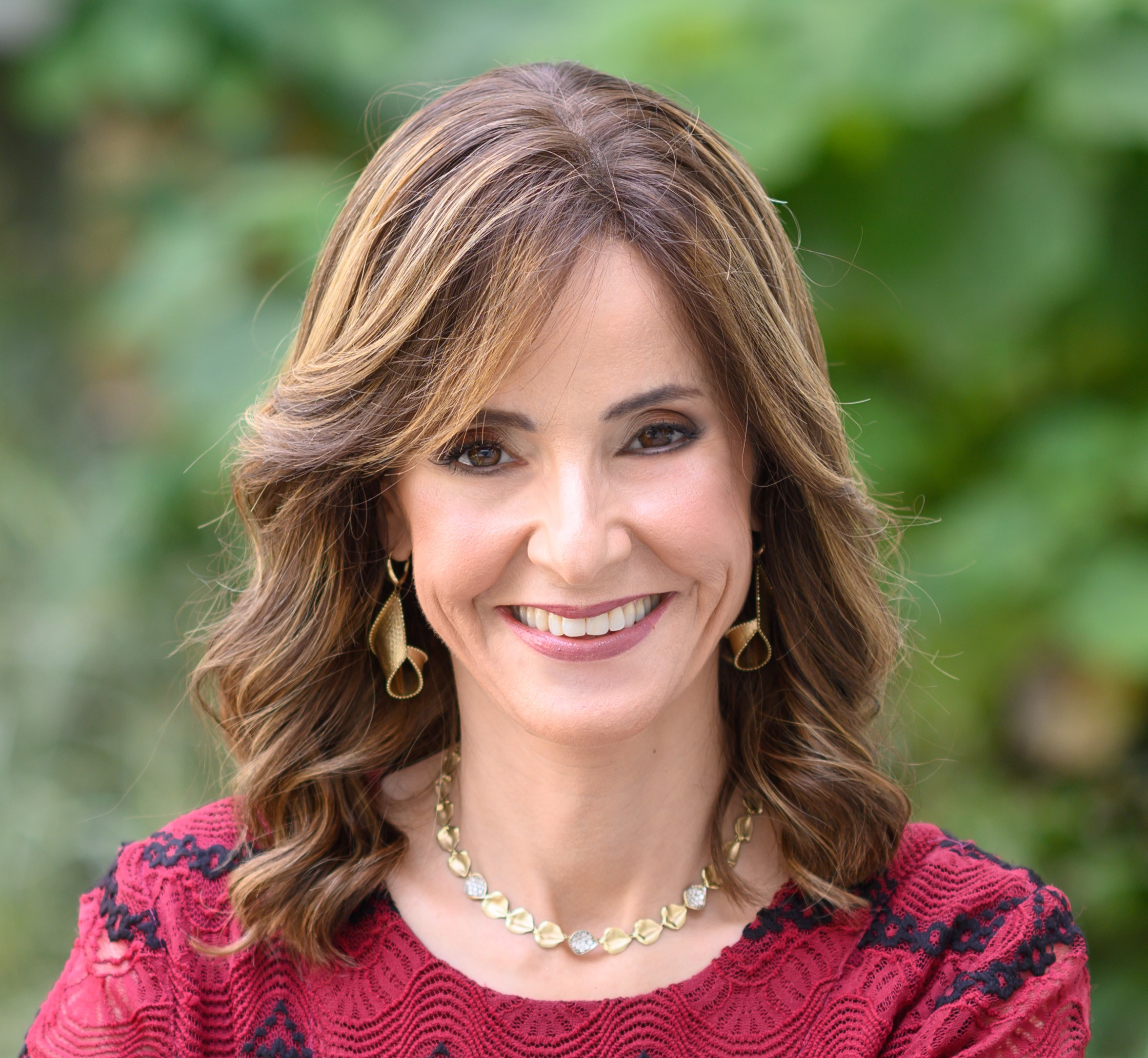}}]{Yonina C. Eldar}
is a Professor in the Department of Mathematics and Computer Science, Weizmann Institute of Science, Rehovot, Israel where she heads the center for Biomedical Engineering and Signal Processing and holds the Dorothy and Patrick Gorman Professorial Chair. She is also a Visiting Professor at MIT, a Visiting Scientist at the Broad Institute, and an Adjunct Professor at Duke University and was a Visiting Professor at Stanford.  She is a member of the Israel Academy of Sciences and Humanities, an IEEE Fellow and a EURASIP Fellow. She received the B.Sc. degree in physics and the B.Sc. degree in electrical engineering from Tel-Aviv University, and the Ph.D. degree in electrical engineering and computer science from MIT, in 2002. She has received many awards for excellence in research and teaching, including the IEEE Signal Processing Society Technical Achievement Award (2013), the IEEE/AESS Fred Nathanson Memorial Radar Award (2014) and the IEEE Kiyo Tomiyasu Award (2016). She was a Horev Fellow of the Leaders in Science and Technology program at the Technion and an Alon Fellow. She received the Michael Bruno Memorial Award from the Rothschild Foundation, the Weizmann Prize for Exact Sciences, the Wolf Foundation Krill Prize for Excellence in Scientific Research, the Henry Taub Prize for Excellence in Research (twice), the Hershel Rich Innovation Award (three times), and the Award for Women with Distinguished Contributions. She received several best paper awards and best demo awards together with her research students and colleagues, was selected as one of the 50 most influential women in Israel, and was a member of the Israel Committee for Higher Education. She is the Editor in Chief of Foundations and Trends in Signal Processing, a member of several IEEE Technical Committees and Award Committees, and heads the Committee for Promoting Gender Fairness in Higher Education Institutions in Israel.
\end{IEEEbiography}
\vskip -1\baselineskip plus -1fil 

\begin{IEEEbiography}[{\includegraphics[width=1.1in,height=1.25in,clip,keepaspectratio]{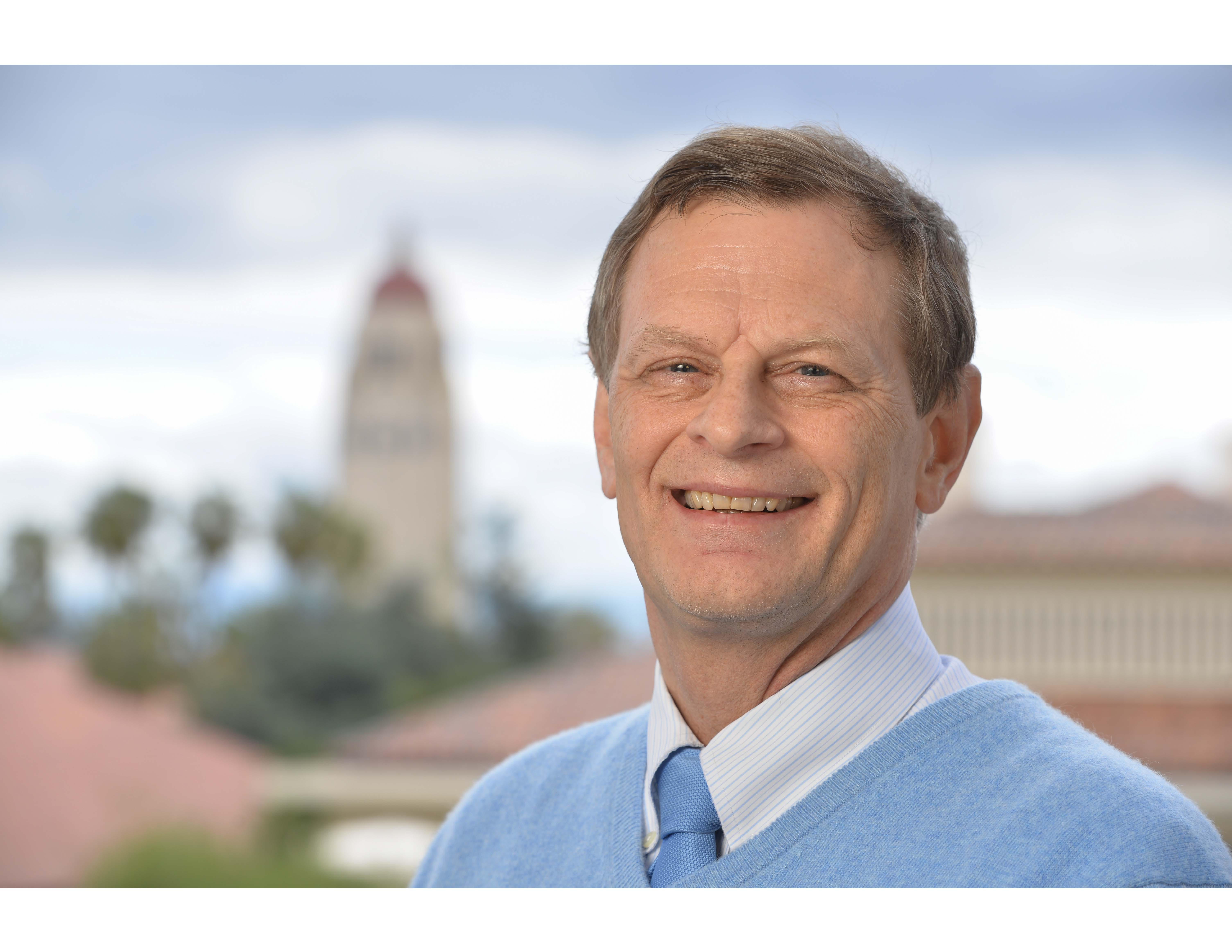}}]{John M. Cioffi} -- Illinois-BSEE: 1978, Stanford-PhDEE: 1984; Prof. EE, Stanford, 1986-present, now recalled emeritus.  Bell Labs 1978-84; IBM Research 84-86.  Founder Amati 1991 (1997 purchased by TI); Chairman and CEO ASSIA Inc.  Cioffi's specific interests are in the area of high-performance digital transmission. Awards include USA's Medal of Technology (2023), IEEE AG Bell (2010), Kirchmayer (2014) and Millennium Medals; Member Internet (2014) and Consumer-Electronics (2018) Halls of Fame; Marconi Fellow (2006); Member, US National (2001), UK Royal (2009), and Scottish Royal (2024) Engineering Academies; IEEE Kobayashi (2001), Armstrong (2014), and Women-In-Coms-Mentor (2018) Awards.  Cioffi has served over a dozen boards of directors, presently PhyTunes (Chairman) and Marconi Society, along with Advisory Boards of VectorML and Zenith Aerospace. 800+ papers and 150+ heavily licensed patents.
\end{IEEEbiography}

\end{document}